\documentclass{article}

\usepackage{arxiv}
\usepackage[utf8]{inputenc}
\usepackage[T1]{fontenc}
\usepackage{hyperref}
\usepackage{url}
\usepackage{booktabs}
\usepackage{amsfonts}
\usepackage{nicefrac}
\usepackage{microtype}
\usepackage{graphicx}
\usepackage{doi}
\usepackage{subcaption}
\usepackage{amssymb, amsfonts, amstext, amsmath}
\usepackage[english, ruled, vlined]{algorithm2e}

\title{AI-Powered Dynamic Fault Detection and\\Performance Assessment in Photovoltaic Systems}

\author{
\href{https://orcid.org/0009-0003-7636-4150}{\includegraphics{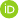}\hspace{1mm}Nelson Salazar-Peña}\\
Department of Mechanical Engineering\\
Universidad de los Andes\\
Bogotá, Colombia \\
\texttt{na.salazar10@uniandes.edu.co} \\
\And
\href{https://orcid.org/0000-0002-6630-5582}{\includegraphics{orcid.pdf}\hspace{1mm}Alejandra Tabares}\\
Department of Industrial Engineering\\
Universidad de los Andes\\
Bogotá, Colombia \\
\texttt{a.tabaresp@uniandes.edu.co} \\
\And
\href{https://orcid.org/0000-0002-0663-2653}{\includegraphics{orcid.pdf}\hspace{1mm}Andrés González-Mancera\thanks{Corresponding author}}\\
Department of Mechanical Engineering\\
Universidad de los Andes\\
Bogotá, Colombia \\
\texttt{angonzal@uniandes.edu.co} \\
}

\renewcommand{\shorttitle}{\textsc{AI-Powered Dynamic Fault Detection and\\Performance Assessment in Photovoltaic Systems}}

\hypersetup{
pdftitle={AI-Powered Dynamic Fault Detection and Performance Assessment in Photovoltaic Systems},
pdfauthor={Nelson Salazar-Peña, Alejandra Tabares, Andrés González-Mancera},
pdfkeywords={First keyword, Second keyword, More},
}

\begin{document}
\maketitle

\begin{abstract} \label{sec:abstract}
The primary limitation of photovoltaic (PV) solar energy lies in its intermittent nature, driven by variable weather conditions, which makes it susceptible to significant power losses and failures. Research indicates that power losses can range from 10 to 70\%, with a mean decrease in energy production of 25\% \cite{ref:C1R10}.

A precise characterization of losses, along with effective fault detection, is crucial for reliable performance assessment of PV systems. This not only aids in making informed decisions aimed at achieving the desired PV system efficiency but also underscores the importance of integrating this information into the control signal monitoring system alongside production parameters \cite{ref:C1R14}.

Computational modeling of PV systems offers a robust framework for conducting technological, economic, and performance analyses. The traceability of algorithms used in these models ensures their acceptance within both technical and financial sectors. However, current computational models are often rigid, limiting their ability to perform advanced performance analyses necessary for optimizing PV system performance and driving technological innovation \cite{ref:C2R1}.

Conventional fault detection strategies for PV systems often fall short due to the complexity of the data signal profiles they must analyze. These methods frequently involve additional costs related to equipment and personnel, extended execution periods, potential interruptions in energy production, and, in many cases, unreliable results.

In contrast, artificial intelligence (AI) methods have gained traction, particularly through machine learning algorithms trained to understand the relationship between input parameters (e.g., meteorological and electrical variables) and output parameters (e.g., production metrics). Once a well-trained model is developed, it can effectively identify faulty states in the PV system by evaluating deviations from expected performance \cite{ref:C3R2, ref:C3R3}.

This research presents a computational model designed to evaluate the performance of a PV system, with a built-in fault detection mechanism. The model, developed using the PVlib library in Python, includes a dynamic loss quantification algorithm that assesses meteorological, operational, and technical data from the PV system. The fault detection component is based on an artificial neural network (ANN) trained on synthetic datasets with a five-minute resolution, reflecting dynamic climatic conditions and associated physical characteristics. These datasets also incorporate randomly occurring faults to simulate the stochastic nature of faults in real-world PV system operations. Additionally, a dynamic threshold definition strategy for fault detection is developed based on historical meteorological data and operational information from a PV system located at Universidad de los Andes.

The key contributions of this research are as follows: (i) a computational model of a PV system with a minimum mean absolute error of 6.0\% in daily energy estimation; (ii) a dynamic quantification of losses that accounts for climatic conditions and system production levels without the need for specialized meteorological equipment; (iii) an AI-based algorithm for estimating the time signals of technical parameters from system equipment, eliminating the need for special monitoring devices; and (iv) a fault detection model that operates without requiring detailed technical information about the equipment or PV system topology, achieving a mean fault detection accuracy of 82.2\% and a maximum accuracy of 92.6\%.

\end{abstract}

\keywords{Photovoltaic solar energy \and Fault detection \and Computational modeling \and Artificial intelligence \and Loss quantification} \label{sec:keywords}

\section{Introduction}
\label{sec-1}

\subsection{Motivation}
\label{subsec:C1S1S1}
Collective awareness regarding environmental issues is fostering initiatives that promote energy sustainability. For this reason, the energy industry is shifting from the use of non-renewable fossil resources to clean sources that involve minimal environmental impact, both in their generation and subsequent use. This shift in thinking is reinforced by observing the evolution of global energy consumption by generation source, where renewable energies have the highest growth rate (12\%, approximately) compared to traditional sources \cite{ref:C1R1}. Furthermore, \cite{ref:C1R8} validates this argument by stating that photovoltaic (PV) solar energy has positioned itself in the last 15 years as the most used renewable energy, also highlighting the fact that PV solar projects now offer some of the cheapest sources of electricity in history \cite{ref:C1R4}. 

\subsection{Problem Statement}
\label{subsec:C1S1S2}
The main weakness of PV solar energy is its intermittent nature due to variable weather conditions (e.g., effective solar irradiance, geographical factors, daily weather conditions, cloud cover, fog or mist, turbidity) and susceptibility to losses and failures (e.g., soiling, shading, degradation).

Losses in a PV system refer to the reduction in energy generation due to external factors that cause a disturbance in the nominal behavior of electrical components (i.e., balance of system (BOS)) in a known range without affecting their integrity. Research in the area estimate that power losses vary between 10 and 70\%, representing an average decrease of 25\% in produced energy \cite{ref:C1R10}.

The following loss factors are highlighted in \cite{ref:C1R9}: air pollution, soiling, shading (soft and hard), angle of incidence, low irradiance, PV panel temperature, PV panels mismatch, degradation, inverter performance, maximum power point tracking (MPPT) performance, tracking reduction, transformer performance, DC and AC cables, auxiliary power source, downtime (blackouts), and grid availability and compliance.

Failures in a PV system, unlike losses, do imply a detriment to the integrity of the BOS; their detection is a key factor to increase the efficiency, reliability, and lifespan of the PV system \cite{ref:C1R11}. Research have estimated the ranges in which energy generation has historically been reduced due to failures. However, given the danger they can represent (e.g., electric discharges or fires), it is common to cause a system blackout \cite{ref:C1R12}. The main failures documented in the literature \cite{ref:C1R11, ref:C1R12, ref:C1R13} are line-to-line, line-to-ground, bypass diode, junction box, open circuit, electric arc, hot spot, delamination, discoloration, and microcracks.

For investors, owners, and academia, the importance of knowing the losses and failures precisely lies in \cite{ref:C1R9}:

\begin{itemize}
    \item \textbf{Investors:} To evaluate the project's viability and provide financing, it is necessary to predict the average annual energy production during the lifespan of the PV system (generally 25 to 30 years). The estimation of energy yield allows approximating the project's revenue.
    
    \item \textbf{Owners:} To evaluate of the PV system's performance to know its condition. Metrics and indicators, such as performance ratio, fill factor, and capacity factor, are used for this purpose.
    
    \item \textbf{Academia:} To establish a reference framework based on research (e.g., loss estimates). Approaches such as simulation studies, production data analysis, and evaluation of meteorological and spatial conditions are common.
\end{itemize}

Undoubtedly, an accurate characterization of losses and fault detection consolidates a reliable performance evaluation of PV systems. In turn, it allows guiding decision-making towards achieving the expected PV system efficiency \cite{ref:C1R14}. Therefore, it is increasingly necessary to have this information as an implementation to the control signal monitoring system, together with the production parameters. 

\subsection{Novelty and Contributions}
\label{subsec:C1S2S4}
This research presents a novel computational model for the performance evaluation and fault detection of PV systems, with application to the PV system at Universidad de los Andes. The primary contributions and novel aspects of this work are:

\begin{enumerate}
    \item \textbf{High-precision computational modeling:} Developed a computational model with high accuracy in operational parameter estimation, achieving a mean absolute error of 6\% for simulated energy.

    \item \textbf{Dynamic loss quantification:} Implemented a dynamic quantification approach for various loss factors including soiling, degradation, DC and AC wiring, and inverter performance, without the need for specialized meteorological equipment, and also adjusts losses based on the PV system's production level and considering the PV system’s startup thresholds and saturation limits.

    \item \textbf{Artificial intelligence (AI)-powered fault detection:} Devised an AI algorithm capable of estimating technical parameters without the need for specialized monitoring devices nor technical specificacions of the PV system. The algorithm demonstrates strong performance with mean absolute error of 9.6\%.

    \item \textbf{Establishment of normal operating thresholds:} Developed a methodology for defining normal operating thresholds based on statistical metrics derived from historical production data and climatic dynamics. This approach addresses the limitations of conventional methods that often fail to account for dynamic meteorological conditions.

    \item \textbf{Robust fault detection algorithm:} Implemented a fault detection algorithm using an AI model and statistical descriptors that does not rely on detailed technical information about the equipment or system topology. The algorithm achieves high accuracy in fault detection with a mean accuracy of 82.2\% and a max accuracy of rates of 92.6\%.
\end{enumerate}

\subsection{Structure}
\label{subsec:C1S3}
This paper is organized into three main stages that detail the development and validation of a computational model for PV systems and its application in fault detection. Additionally, a detailed literature review is included (Section \ref{sec-3}) on monitoring systems, fault detection techniques, and computational modeling for the deployment and validation of simulations. The structure of the paper is outlined as follows:

The first stage focuses on developing the computational model (Section \ref{sec-4}). It begins with a comprehensive literature review (Section \ref{subsec:C2S2S1}) of existing models that simulate PV system performance, aiming to understand the application of computational methods across various research areas. This review also examines the key variables and functions that influence the behavior of PV panels and the BOS components, and considers the assumptions inherent in different computational approaches.

The selection criteria for the model, methods, and libraries (Section \ref{subsec:C2S2S2}) include: (i) the purpose of the model; (ii) the programming language used; (iii) access to the backend; (iv) technical rigor; (v) construction flexibility; (vi) validation precision; (vii) economic considerations; and (viii) repeatability and reproducibility.

Subsequently, the computational model is constructed based on the PV system at Universidad de los Andes. To enhance simulation accuracy, loss factors affecting the PV system are estimated (Section \ref{subsec:C2S2S3}). This involves a literature review of loss factors and evaluation techniques, followed by the identification and implementation of relevant algorithms and methods into the model. Validation of the model is carried out by comparing simulation estimates with actual measurements from a monitoring system using statistical metrics.

The second stage centers on developing and assessing fault detection algorithms (Section \ref{subsec:C3S2S4} to \ref{subsec:C4S2S2}). This stage uses synthetic data generated from real PV system information, allowing precise control over the state of the system (normal or faulty). Synthetic data generation is detailed in Section \ref{subsec:C3S2S1} to \ref{subsec:C3S2S3}, including the incorporation of known faults. A literature review identifies common fault factors and their impact on energy production, which are then adapted to the specific characteristics of the study plant (e.g., configuration, location, BOS).

The final stage involves integrating the computational model with the fault detection system (Section \ref{sec:C2S3} to \ref{subsec:C3S3S3}). This integration is validated through computational simulations that analyze synthetic daily production data with potential faults (Section \ref{subsec:C3S3S4}). The accuracy, robustness, and effectiveness of the combined model are evaluated using statistical metrics to ensure alignment with the research objectives (Section \ref{sec:C4S3}).

Each stage builds upon the previous one, culminating in a comprehensive evaluation of the computational model's performance and its capability for fault detection in PV systems. The results obtained are finally contrasted with those presented in the literature (Section \ref{subsec:C4S3S6}).

Finally, the conclusion and future work of the research are presented in Section \ref{sec-7} and \ref{sec-8}, respectively.

\subsection{PV System}
\label{subsec:C1S3S2}
In 2019 the PV system at Universidad de los Andes began operation. The system has an installed capacity of 80.1 kW connected to the grid (on-grid) and consists of 200 PV panels distributed between two central inverters (referred to as \emph{System A} and \emph{System B} hereinafter). The PV system is equipped with a monitoring system developed by Meteocontrol. Irradiance monitoring is carried out using the Kipp \& Zonen CMP3 pyranometer and the Si-I-420-T reference cell.  These are
Class A devices according to \cite{ref:C1R16}. Lastly, data acquisition for the monitoring system is done through the Data Logger blue’Log XM-200 device, allowing data recording with temporal resolutions of five minutes, thirty minutes, hourly, daily, and monthly. Additional technical information is provided in Table \ref{tab:C1T5}.

\begin{figure}[]
\centering
\includegraphics[width=0.65\textwidth]{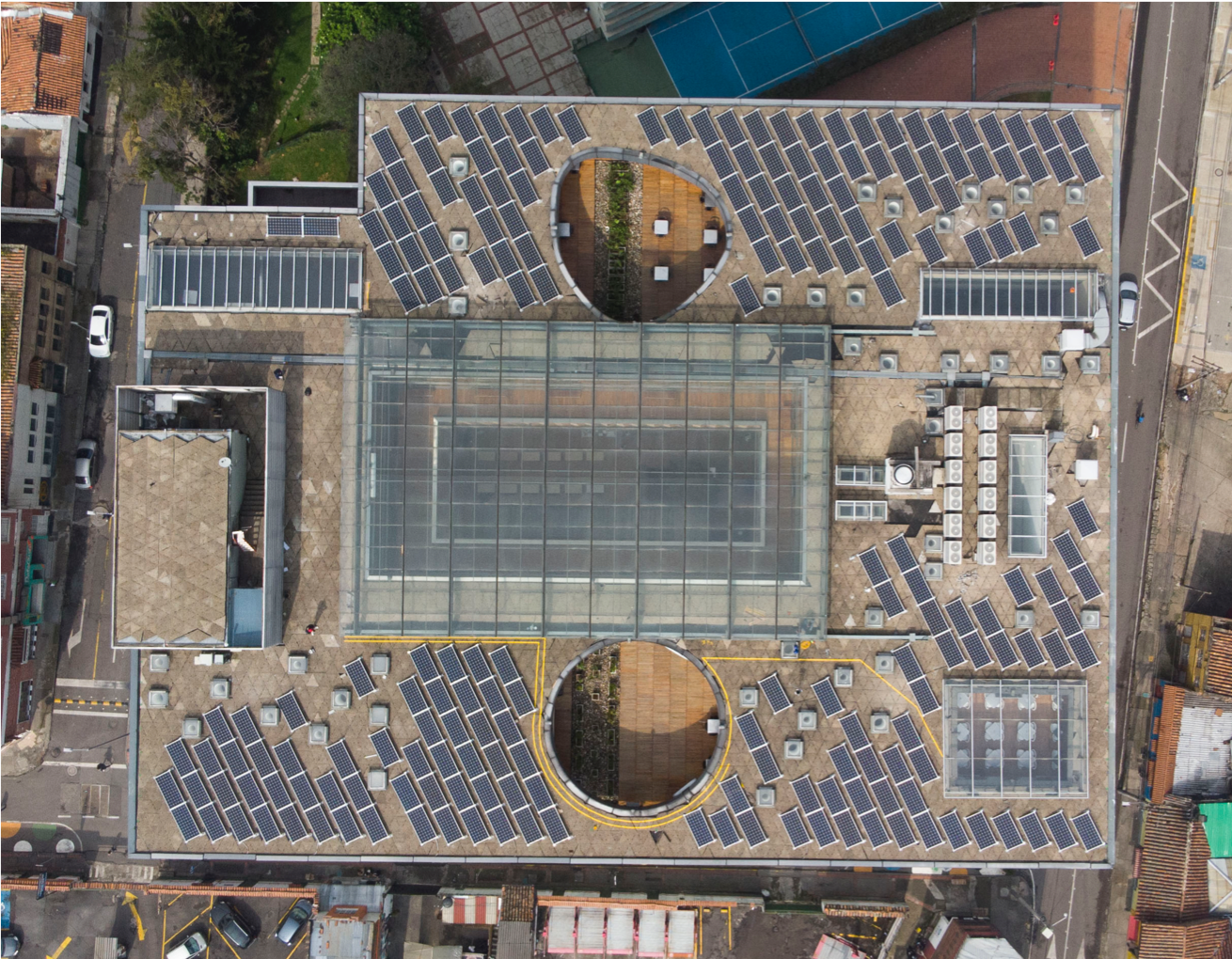}
\caption{PV system at Universidad de los Andes. Taken from \cite{ref:C1R62}.}
\label{fig:C1F4}
\end{figure}

\begin{table}[]
\centering
\caption{Technical information of the PV system at Universidad de los Andes.}
\label{tab:C1T5}
\begin{tabular}{ccc}
\hline
Technical Spec & System A & System B \\ \hline
AC Power (kW) & 51.24 & 28.82 \\
Tilt ($^{\circ}$) & 10 & 10 \\
Azimuthal ($^{\circ}$) & 180 & 180 \\
PV Panels per String & 16 & 18 \\
Strings per Inverter & 8 & 4 \\
Inverter & ABB TRIO 50.0 TL OUTD US & ABB TRIO 27.6 TL OUTD US \\
PV Panel & LG Electronics LG400N2W-A5 & LG Electronics LG400N2W-A5 \\ \hline
\end{tabular}%
\end{table}

\subsection{Computational Tools}
\label{subsec:C1S3S3}
The standard computational development is carried out on a macOS operating system with a 2.7 GHz Intel Core i5 dual-core processor and 8 GB RAM. Advanced processes requiring greater computational capacity are performed on a specialized computer with a sixteen-core Xeon E5 2630 v3 processor, 96 GB RAM, GTX 1080 Ti graphics card, Linux operating system with amd64 architecture, Ubuntu 18.04 distribution, and equipped with CUDA 10.2.

\section{Literature Review}
\label{sec-3}

\subsection{Monitoring Systems}
\label{subsec:C1S3S1}
Monitoring systems are supervisory tools whose function is to collect information about the meteorological and operational parameters of the PV system with the aim of evaluating its performance and providing maintenance and future improvements \cite{ref:C1R15}. 

In \cite{ref:C1R16} the guidelines for a proper monitoring of PV system performance are outlined. It also highlights the following purposes of a monitoring system: (i) identify performance trends; (ii) locate system faults; (iii) compare system performance against design expectations and warranties; (iv) compare systems with different configurations and in different locations; and (v) analyze performance trends and fault localization.

For instance, \cite{ref:C1R18, ref:C1R19} uses a set of equations to extrapolate point-to-point energy generated to energy at standard test conditions (STC). This facilitates the interpretation of the monitored signal and, therefore, the diagnosis and detection of instantaneous decreases in generated power; its utility is highlighted when the variation of power with temperature is not known. 

Different from the representation in STC conditions, the voltage-current curve (I-V curve) of a PV panel provides information related to the electrical behavior (i.e., normal or faulty). It is the most important and widely used metric to describe the electrical output of a PV panel \cite{ref:C1R20}. For example, \cite{ref:C1R21} characterizes the behavior of operating parameters of PV panels and verifies the presence of faults by contrasting the expected power against the obtained power from the I-V curve. Similarly, \cite{ref:C1R22} propose another strategy based on the I-V curve; their method obtains the statistical distribution of each electrical parameter of the I-V curve and determines the probability density function that best describes the dispersion of these parameters. By comparing the statistical results against the expected maximums, mismatch losses are estimated.

Another approach focuses on developing performance metrics to expand the characterization of the production parameters of the PV system. \cite{ref:C1R18, ref:C1R19} highlights metrics derived from the installed capacity of the PV system (e.g., performance ratio) and based on a more detailed model of PV system performance (e.g., power performance index).

The performance ratio compares the measured production and the value of the installed capacity \cite{ref:C1R9}. Its usefulness lies in providing a benchmark for comparing PV systems over a specific period \cite{ref:C1R16}. A higher value indicates a more efficient PV system in converting solar irradiance into useful energy. Additionally, it quantifies the overall effect of system losses on the nominal capacity, including losses caused by low irradiance, temperature, PV panels, mismatch, inverters, wiring, shading, and soiling \cite{ref:C1R16}. Similarly, the capacity factor parameter is the ratio between the measured production during a year and its production if it had operated at nominal power in the same period \cite{ref:C1R16}.

Regarding metrics based on detailed models of PV system performance, \cite{ref:C1R16, ref:C1R9} establishes that they are based on comparing actual production against predicted or expected. For example, \cite{ref:C1R24} develops a metric that relates the measured power to the expected power, with the particularity of normalizing for irradiance, temperature, and other power loss mechanisms. Thus, this metric allows knowing the actual performance of the PV system, without biases from losses or seasonal variations. 

However, the most relevant metric is the ill factor, which is obtained from the I-V curve. According to \cite{ref:C1R21}, the fill factor describes the slope of the I-V curve and also the relationship between the ideal power (obtained from the multiplication of the open-circuit voltage $V_{oc}$ and short-circuit current $I_{sc}$) and the power at the maximum power point (MPP) $P_{DC}$.

For instance, \cite{ref:C1R25} explores DC production signals and their residual signal. The time correlation of these signals is statistically characterized using an autoregressive model that, when exceeding a threshold, indicates the occurrence of a fault. The model has a normalized root mean squared error of 2\%. Similarly, \cite{ref:C1R26} rely on the variation of voltage and current from the I-V curve for fault detection. The detection principle involves comparing the signal deviation of each PV panel with respect to the string. If a particular signal has a significant deviation, a fault state for that PV panel is asserted. 

In more recent research, \cite{ref:C1R27} studies the geometry of the I-V curve to diagnose shading losses and faults due to cracks in PV panels. The method involves deriving the I-V curve to assess convex behavior in its geometry. In the convex range, if the difference between the current of the derivative curve and the I-V curve is greater than a certain threshold, the occurrence of shading loss or cracking fault in the PV panel is detected. On the other hand, \cite{ref:C1R28} manages to detect line-to-line and line-to-ground faults through a Fourier analysis of the DC current signal, taking advantage of the fact that when a fault occurs in a DC system, the current acquires an oscillatory behavior, adding additional components to the signal. Fault detection occurs when the ratio between the expected current magnitude and that obtained from the main harmonics of the Fourier analysis exceeds a given threshold. Detection is achieved within 0.5 milliseconds of the fault onset. Finally, \cite{ref:C1R29} relies on the estimation of the percentage of power loss from the power-voltage curve to detect hot spot faults, based on the ratio between the measured power of the PV panel affected by the hot spot and the mean power of adjacent PV panels. Detection is performed by correlating the cumulative distribution function (CDF) with the CDFs of known hot spot faults. The accuracy of this strategy is 80\%.

\subsubsection{Techniques and Methods}
\label{subsec:C1S3S2}
There are various techniques and methods that allow the identification and localization of faults in PV systems. These tools must be robust to provide safe operation, increase system reliability and lifespan, be quick in fault localization to restrict major consequences, and capable of detecting multiple faults without interfering with production \cite{ref:C1R11, ref:C1R31, ref:C1R30}.

The techniques and methods are grouped into six categories: (i) visual methods; (ii) imaging solutions; (iii) electrical methods; (iv) data analysis techniques; (v) electrical protection devices; and (vi) arc fault detectors \cite{ref:C1R11, ref:C1R13, ref:C1R32}.

Visual methods involve optical inspections of PV panels performed regularly by trained personnel. According to \cite{ref:C1R11, ref:C1R32, ref:C1R33}, visual inspection should be conducted at 1 000 lux and from different angles to avoid reflections. The defects in PV panels that can be detected with this technique include: discoloration, bubbles, delamination, browning, fractures, microcracks, soiling, wiring deterioration, and oxidation. However, the main drawback of the visual method is its dependence on human capabilities, which can lead to delays in detecting faults and yield unreliable results. Recently, as mentioned by \cite{ref:C1R32}, unmanned aerial vehicles are used to mitigate these issues as they are equipped with sensors and onboard digital cameras and  provide images and data to the control unit for enhanced fault detection. 

Imaging solutions employ equipment and tools to obtain a qualitative characterization of the PV panel. Information from the images is extracted, and with the aid of image processing techniques, the location of faults is identified \cite{ref:C1R32}. The main strategies include:

\begin{itemize}
    \item \textbf{Infrared thermography}. It is based on estimating temperatures by capturing the infrared radiation from the electromagnetic spectrum. In the interconnection of PV panel cells, certain cells generate less current than others due to degradation mechanisms or mismatch, leading to increased resistance and heat dissipation (i.e., hot spot), observed as a bright point in thermal images \cite{ref:C1R32}. This strategy also allows the identification of loose connections and increased series resistance \cite{ref:C1R12}.
    
    \item \textbf{Electroluminescence}.
    It involves applying voltage to the PV panel to induce electroluminescence and reveal subsurface defects of a PV cell, such as fractures, microcracks, or non-uniform current \cite{ref:C1R11, ref:C1R12, ref:C1R32}. The main drawbacks are its high cost and the requirement for offline testing, as well as challenges in crack identification due to varied texture backgrounds, low contrast between cracks and the surrounding background. To address this, \cite{ref:C1R32} propose an automatic crack detection technique with an average detection rate of 94.4\%, involving the generation of a crack prominence map to improve contrast. 
    
    \item \textbf{Lock-in thermography}. It verifies power loss in a PV cell by injecting current pulses using a lock-in excitation device. The injected current increases the PV cell's temperature, revealing defects. Changing the modulation of the injected current allows detection of different types of defects. This strategy is suitable for detecting small defects but is limited to offline testing \cite{ref:C1R11, ref:C1R12}. 
\end{itemize}

Electrical methods perform fault detection based on the characterization of electrical production parameters. These methods require specific measurement data such as electrical variations, irradiance, and meteorological factors, along with data derived from I–V curve measurements \cite{ref:C1R32}. According to \cite{ref:C1R11, ref:C1R12}, the main strategies are:

\begin{itemize}
    \item \textbf{Climate-independent detection}. This strategy does not involve climatic effects and is based on the analysis of induced signal waveforms \cite{ref:C1R11, ref:C1R13}. Fault detection involves injecting a signal using a signal generator and analyzing the PV system's response using external devices. Two notable methods are: time domain reflectometry, where the delay between the injected and reflected signals allows detecting faults (e.g., an increase in series resistance between PV panels \cite{ref:C1R11, ref:C1R12}), and earth capacitance measurement, which detects disconnection of the PV panel in a string \cite{ref:C1R12}. 
    
    \item \textbf{I-V curve measurement}. It involves evaluating disturbances in the I-V curve to estimate faults affecting production \cite{ref:C1R11}. For example, \cite{ref:C1R34} presents an automatic detection method based on analyzing the static characteristics of the I-V curve, which is divided into voltage and current zones. Fault detection occurs by observing deviations from monitored I-V curves. Similarly, \cite{ref:C1R36} evaluates the relationship of operation parameters from the I-V curve with the expected optimal curve to detect faults. Fault diagnosis occurs by evaluating deformations in the I-V curve; if these deformations exceed a certain threshold, the fault is detected. Finally, \cite{ref:C1R37} proposes the first and second derivatives of the I-V curve to detect series resistance and bypass diode faults. 
    
    \item \textbf{Power loss analysis}. The analysis involves comparing measured results with those obtained from simulating the PV system's behavior. For example, \cite{ref:C1R38} detects faults when the comparison exceeds a 5\% discrepancy, and \cite{ref:C1R40} proposes analyzing anomalies in the characteristics of the PV panel string terminals. The voltage difference between faulty and healthy PV panels allows detecting and locating anomalies. 
   
   \item \textbf{Comparison of measurement-modeling}. It involves comparing measured and expected production obtained through simulation to detect different types of faults in the PV system \cite{ref:C1R11}. For example, \cite{ref:C1R41} proposes estimating the absolute power ratio error from the difference between the power ratio predicted by simulation and the measured value to detect faults. Similarly, \cite{ref:C1R43} analyzes the difference between measurement and simulation using the exponentially weighted moving average to detect faults. MPP voltage and current residuals allow distinguishing between open circuit, short circuit, and shading faults. 
\end{itemize}

As previously discussed, PV systems often face electrical faults due to anomalies in the internal system configuration. Failures such as line-line and line-ground faults are among the most common, and they can be dangerous due to the electrical discharge involved, potentially leading to fires. Therefore, the presence of electrical protection devices and arc fault detectors in PV systems is crucial as they instantly interrupt the electrical circuit to stop ongoing faults \cite{ref:C1R11, ref:C1R44}. According to \cite{ref:C1R30, ref:C1R31, ref:C1R44}, PV systems must be equipped with overcurrent protection devices, ground fault detection and interruption, and arc fault circuit interrupter to protect against line-line, line-ground, and arc faults, respectively. The drawbacks relies on the following limitations: (i) the current discharge may not exceed the fuse threshold and remain undetected; (ii) differential current measurements may contain external noise and trigger a shutdown of the PV system; and (iii) the presence of a MPPT significantly reduces the magnitude of the fault current, and therefore, faults may go undetected.

To overcome these challenges, \cite{ref:C1R45} propose a residual current device which can detect the difference in input and output currents through different PV panels and isolate the specific string in case of fault detection. Other alternatives involve offline measurement of insulation resistance using insulation monitoring devices to detect ground faults; detection occurs if the measured resistance is lower than the preset value \cite{ref:C1R40, ref:C1R45}.  

Regarding arc fault circuit interrupters, these analyze the voltage and current signals of a PV system for abnormalities in the output waveform that indicate arc faults \cite{ref:C1R31}. According to \cite{ref:C1R45}, the optimal frequency range to detect arc fault noise is between 1 and 100 kHz. However, according to \cite{ref:C1R31}, the limitations are: (i) installing the arc fault circuit interrupter (AFCI) in the inverter attenuates the arc signal, making detection challenging; (ii) detecting arc faults is achieved through multiple installed AFCIs, which is costly; and (iii) frequencies of power electronic devices in a PV system can interfere with AFCIs and cause false alarms.

\subsubsection{Artificial Intelligence}
\label{subsec:C1S3S3}
In fault detection based on AI, machine learning (ML) algorithms are trained to learn the relationship between the input and output parameters of a PV system. Once an accurately trained model is constructed, it is possible to identify faulty states of the PV system through error evaluations \cite{ref:C1R31, ref:C1R44}. 

Training data for ML algorithms is collected through monitoring systems or generated with the help of computational simulation models. However, anomalous data during fault occurrences are also required for accurate training and prediction \cite{ref:C1R31, ref:C1R44}. 

Despite presenting advantages such as overcoming the difficulty of defining thresholds, any recently introduced condition can be identified based on prior learning, and aiding in accurate fault detection and classification \cite{ref:C1R12, ref:C1R31}. However, according to \cite{ref:C1R31}, the main disadvantages are: (i) accuracy depends on the quality of training data; (ii) collecting training data through monitoring, especially fault occurrences, is non-trivial; (iii) dependence on the accuracy of simulation models for the reliability of training data and the ML algorithm; and (iv) non-availability of open-source data globally due to variations in PV system type, size, configuration, and location.

There are also hybrid models, where two or more ML techniques are combined to strengthen the accuracy of detection results. According to \cite{ref:C1R47}, the most popular ones are radial basis function neural network (RBFNN), adaptive neuro-fuzzy inference system, and artificial neural networks (ANN) coupled with physical models, which consists of a set of equations describing parameter behavior based on physical principles. 

For example, \cite{ref:C1R48} propose an ANN that can identify short circuit faults, open-circuit faults, bypass diode shading, reversed bypass diode, connection fault, partial shading, shading with faulty bypass diode, and shading with connection fault in the PV system. Similarly, \cite{ref:C1R49} integrate ANN with RBFNN to detect faults based on voltage and power variations, achieving an accuracy of 92\%. \cite{ref:C1R50} uses fuzzy logic for fault detection, mainly bypass diode faults, with an accuracy between 90 and 98\%. 

In addition, \cite{ref:C1R51} propose comparing measured voltage and current values with ANN estimates to detect faults. The ANN estimates voltage and current based on the irradiance and and PV panel temperature, with training data obtained from real-time faulty conditions. Detection occurs when the difference exceeds a threshold. On the other hand, \cite{ref:C1R52} develop a method based on theoretical analysis of I-V curves and fuzzy logic for fault detection on the DC side of a grid-connected PV system. The algorithm focuses on detecting faults caused by partial shading of PV panels, with an accuracy exceeding 98\% \cite{ref:C1R53}. 

Finally, \cite{ref:C1R54} use random forest to detect line-line faults, open circuit faults, and faults caused by partial shading of PV panels. Training data include real-time operating voltage and current. The accuracy exceeds 97\%. \cite{ref:C1R55} detect the same faults at the string level using the K-nearest neighbors algorithm. The required data come from the manufacturer's datasheet under STC and nominal operating cell temperature (NOCT). The model focuses on tracking the I-V curve at different irradiance and temperature levels, and achieves a fault classification accuracy of 98\%.

\subsection{Computational Modeling}
\label{sec:3.2}
The importance of modeling an PV system lies in the ability to conduct a technological, economic, and performance analysis. The results obtained from such modeling enable decision-making that is crucial both in the pre-feasibility stage and during the operation of the project, as highlighted in \cite{ref:C2R1, ref:C2R2, ref:C2R3}.

Moreover, information from computational models can be presented in a way that is accepted in technical and financial industries due to the traceability of algorithms and the familiarity of the industry with the results \cite{ref:C2R1}.

However, current computational models lack flexibility in the simulations they can perform, making it challenging to explore advanced topics in PV performance analysis needed for conducting detailed studies ensuring system optimization and technological innovation \cite{ref:C2R1}. Additionally, having incomplete models, such as those with incomplete documentation, creates confusion in configuration aspects and the correct execution of a simulation. Access to detailed information ensures the reliability and accuracy that computational models must have to guarantee the quality of the results.

According to \cite{ref:C2R1}, the main drawbacks of computational models are:

\begin{itemize}
    \item \textbf{Comparison of performance results using different models and assumptions}. Each simulation makes different assumptions and lacks a standard validation process. A standard validation process ensures the comparison between different computational models, their results, and evaluation of strengths and weaknesses according to the application.
    
    \item \textbf{Ability to modify, customize, and update modeling algorithms}. In developing algorithms to model a PV system, user-friendly interfaces and backend accessibility are conflicting attributes. Commercial software often sacrifice backend access to make their software as user-friendly as possible.
    
    \item \textbf{Ability to view intermediate modeling results and perform on-the-fly statistical analysis}. The output of commercial models reports user-specified parameters but does not allow insight into intermediate estimates of the process. Therefore, the user has control over the initial configuration and, in some cases, input data. This challenge aligns with developers' efforts to make the software user-friendly.
\end{itemize}

To overcome the aforementioned challenges, \cite{ref:C2R4} developed a five-parameter model based on technical data provided by manufacturers and semi-empirical correlation equations. This model predicts the I-V curve and, subsequently, the energy production based on the parameters of the PV panel and specified operating conditions. The five-parameter model is based on the equivalent circuit that models the behavior of a PV panel. The model proposed in \cite{ref:C2R4} is preferred for current computational models as it allows estimating the output power under any operating conditions.

Additionally, \cite{ref:C2R5} introduces PVWatts, developed by the National Renewable Energy Laboratory (NREL), to estimate the electricity production of an on-grid PV system. PVWatts is a user-friendly software that conceals a significant portion of the algorithms modeling PV systems, as well as assumptions about the type, configuration, and operation of the system. For instance, \cite{ref:C2R6} leverages PVWatts to quantify the interannual variability of PV systems using typical meteorological year data, predicting the PV system's production with corresponding uncertainty.

On the other hand, \cite{ref:C2R7} utilize PVsyst to estimate the energy production of a 1 MW PV system and assess its feasibility. Similarly, \cite{ref:C2R8} design and simulate a 1 kW PV system to estimate energy generation with a performance ratio of 72\%.

\cite{ref:C2R9} investigate how PV panels behave concerning irradiance and cell temperature to enhance the reliability of energy simulations. The results show a more accurate energy estimation by adjusting the power dependence on irradiance and cell temperature compared to standard modeling in PVsyst \cite{ref:C2R9}.

Furthermore, \cite{ref:C2R10} employ the PVSOL software to design and estimate the production of a PV system for installation on buildings. The study shows that a 234 kW PV system can generate sufficient energy to balance consumption and reduce dependence on the electrical grid. The study includes a financial analysis. \cite{ref:C2R11} also use PVSOL to design and evaluate three configurations of PV panels with trackers (i.e., fixed, single- and dual-axis) connected to the electrical grid, household appliances, a battery system, and to charge an electric vehicle.

\cite{ref:C2R12} design a string of eleven SunPower 230 W PV panels along with an America Sunny Boy 3800TL US inverter and compare simulation results in PVsyst and the PVlib library for the same geographical location. The results show a difference in AC power estimation between 6 and 10\%. The study, besides validating the flexibility of PVlib, confirms its accuracy against commercial software. According to the authors, the modeling flexibility in PVlib allows implementing optimization algorithms to maximize performance, paving the way for expanded research, analysis, and innovation.

\cite{ref:C2R13} leverage the flexibility of PVlib to include the loss factor model algorithms (i.e., simulation with determination of performance coefficients, quantification of instabilities, and fault diagnosis) to adjust modeling, reduce errors even without spectral information, reference cell, or PV panel temperature measurements. The results indicate a more accurate performance analysis, as well as pointing out abnormalities.

\subsection{Failure Detection}
\label{sec:C3S1}
Conventional strategies for detecting faults in PV systems often prove insufficient due to the complexity of the data signal profiles they handle. These methods also require additional costs in equipment and personnel, long execution periods, potential disruption of energy production, and frequently yield unreliable results.

Methods employing AI, specifically ML techniques, have gained popularity for their ability to surpass the challenge of defining thresholds, identify any newly introduced conditions based on prior learning, and assist in precise fault detection and classification \cite{ref:C3R1, ref:C3R2}.

For instance, \cite{ref:C3R4} propose an automatic fault classification method for PV systems using Convolutional Neural Networks (CNNs) for segmentation and classification from RGB images. The classification of fault and no-fault achieves an average accuracy of 75\%, while the classification among no-fault, cracks, shading, and soiling achieves 70\% accuracy. Similarly, \cite{ref:C3R5} employ CNNs for fault classification in integrated distributed networks with distributed generators. The author performs 10-fold cross-validation to evaluate performance metrics such as sensitivity, specificity, accuracy, and F1-score. The results indicate an average accuracy of 99\%.

Furthermore, \cite{ref:C3R6} propose fault detection and diagnosis based on probabilistic neural networks. The method evaluates three operating scenarios: healthy system, three PV panels of a string in short circuit, and ten PV panels of one string in short circuit and another string disconnected. The results show an 82\% accuracy in detection and 98\% in diagnosis. Moreover, this metric is compared with an ANN, demonstrating an average accuracy approximately 23 percentage points higher.

Regarding ANN, \cite{ref:C3R7} employs them to detect short circuit, diode bypass shunt, and PV panel-to-panel connection failure. The required input data include voltage, current, and the number of peaks in the I-V curve. Fault detection occurs by comparing simulation results against in-situ measurements. Data analysis algorithms are then used to classify detected faults. The results indicate an average accuracy of 90\% for detecting the four types of faults.

In contrast, \cite{ref:C3R8} present two classifiers based on decision trees. Both methods classify faults by extracting and selecting features from operational voltage and current data. The classifiers achieve accuracies of 97\% and 99\%, respectively.

\subsection{Computational Simulation Validation}
\label{sec:C4S1}
For fault detection, AI strategies require comparing predictions with corresponding target values. Direct comparison of data is not sufficient for accurate fault detection due to the complex behavior of PV system signals, such as voltage, current, and power.

For example, \cite{ref:C3R1} propose incorporating a tolerance factor in direct comparisons to avoid false alarms. The same author also emphasizes setting threshold limits to detect faults through computational modeling or theoretical analysis. Similarly, \cite{ref:C3R3} analyze residual signals between prediction and measurement to determine if the system is experiencing constant energy loss conditions. Fault detection thresholds are established by simulating capture losses under clear sky conditions. \cite{ref:C3R1} compare the performance ratio obtained through computational electrical simulation with measurements. Faults are identified when the absolute difference between these two values exceeds 0.07. The threshold value was determined through experimental testing \cite{ref:C3R4}. \cite{ref:C3R5} employ descriptive statistics to compare voltage and current signals estimated by a neural network with measurements from the monitoring system. The coefficient of determination \( R^2 \) is used to establish correlations with shading faults. For instance, an \( R^2 \geq 0.90 \) indicates shading over five cells, whereas \( R^2 \leq 0.50 \) suggests shading over fifty cells.

Finally, \cite{ref:C3R6} establish operational limits based on the relationship between measured AC power and modeled power. The upper and lower thresholds correspond to $\mu \pm 3\sigma$, where $\mu$ is the mean and $\sigma$ is the standard deviation of the values obtained from the relationship. One of the main distinctions of this author's proposal is the correlation between these threshold values and irradiance, where higher fault detection accuracy is achieved by setting thresholds at 200 W/m$^2$ intervals (approximately 96\%). Conversely, when a single threshold is defined for the entire irradiance dataset, the detection accuracy reaches approximately 81\%.

\section{Computational Modeling}
\label{sec-4}

\subsection{Choice of Modeling Tool}
\label{subsec:C2S2S1}
There are numerous commercial and academic algorithms for simulating the performance of PV systems. Although these models aim to estimate energy, they all differ in design, assumptions, conceptual approach, mathematical modeling, and the amount of data required to carry out the simulation \cite{ref:C2R3}.

\cite{ref:C2R14, ref:C2R15} conduct a review of existing models, detailing their capabilities, trade-offs, and deficiencies. Table \ref{tab:C2T1} presents the models that are currently active, along with their purpose, years of development, and programming language.

\begin{table}[]
\centering
\caption{Computational tools for modeling PV systems. Adapted from \cite{ref:C2R14, ref:C2R15}.}
\label{tab:C2T1}
\begin{tabular}{cccl}
\hline
Tool & Available Since & Language & Description \\ \hline
PVlib & 2012 & Matlab, Python & General modeling of PV systems \\
PVWatts & 2004 & C++ & General modeling of PV systems \\
PVsyst & 1995 & C++ & General modeling of PV systems \\
PVSOL & 1998 & C++ & General modeling of PV systems \\
Polysun & 2009 & C++ & General modeling of PV systems \\
Pecos & 2016 & Python & Monitoring the performance of PV systems \\
PVMismatch & 2012 & Python & Estimating I-V curve of cells with mismatch \\
CASSYS & 2015 & Excel, C\# & General modeling of PV systems \\
rdtools & 2017 & Python & Evaluate the degradation of PV systems \\
pvfactors & 2016 & Python & Diffuse shading and bifacial irradiance modeling \\
photovoltaic & 2017 & Python & General modeling of PV systems \\
solaR & 2010 & R & General modeling of PV systems \\
feedinlibg & 2015 & Python & PV time series modelin \\ \hline
\end{tabular}%
\end{table}

The main features of PVWatts are: (i) estimates energy production along with financial analysis; (ii) involves simple modeling with typical values for system type, configuration, and operation; and (iii) the modeling steps are system inputs, solar radiation, module transmittance, thermal model, DC module model, system losses, inverter model, and AC system response.

Regarding PVsyst, its main features include \cite{ref:C2R14}: (i) most widely used commercial modeling application for PV projects; (ii) allows modeling of on-grid, stand-alone, and solar pumping installations; (iii) uses proprietary mathematical models for computational simulation; (iv) provides highly detailed documentation; (v) enables analysis of mismatch in PV panels to determine more specific electrical parameters, and evaluates cell shading and other voltage losses due to wiring and soiling; and (vi) modeling steps include meteorological information, geographic configuration, system design, AC system response, and system losses report.

Nevertheless, the most prominent research tool is PVlib. PVlib is an open-source library in Python that allows for modeling PV systems \cite{ref:C2R15}. It is developed by Sandia National Laboratories (SNL) and continually expands thanks to the photovoltaic performance and modeling collaboration (PVPMC), a group of professionals in the PV field who share information and help improve PV system prediction models \cite{ref:C2R16, ref:C2R17}. The main features of PVlib are \cite{ref:C2R17, ref:C2R18, ref:C2R19}: (i) it leverages the Python programming language, which ensures flexibility and free access for both academic and commercial use; (ii) designed for collaborative development and supported by a rigorous method for incorporating contributions from authors and research into the library; (iii) supported by a set of tests and validations to ensure the library's stability and allow the validation of model results with real performance data; and (iv) allows for modeling and analyzing each part of the PV system production chain, as well as using external data analysis libraries in Python.

Additionally, according to \cite{ref:C2R1}, PVlib promotes the development of the PV industry for the following reasons:

\begin{itemize}
    \item \textbf{Creates a link between researchers and developers}. New algorithms can be presented in a format that quickly adapts to current modeling work.
    
    \item \textbf{Provides validated tools to those who do not have the capacity to develop them}. Industrial and research groups are limited to the capabilities of commercial PV analysis software or manual spreadsheets.
    
    \item \textbf{Collaborative design increases the pace of technological innovation}. The expansion of models consists of contributions from the community, so new methodologies, analysis techniques, and best practices are quickly adopted.
    
    \item \textbf{Facilitates communication with investors.} By providing standard modeling for analyzing advanced data, communication is simplified and shared reproducibly.
\end{itemize}

For these reasons, in this research, the computational modeling of the PV system at Universidad de los Andes is performed using the PVlib library in Python.

\subsection{Construction of the Computational Model}
\label{subsec:C2S2S2}
\begin{figure}[]
\centering
\includegraphics[width=0.8\textwidth]{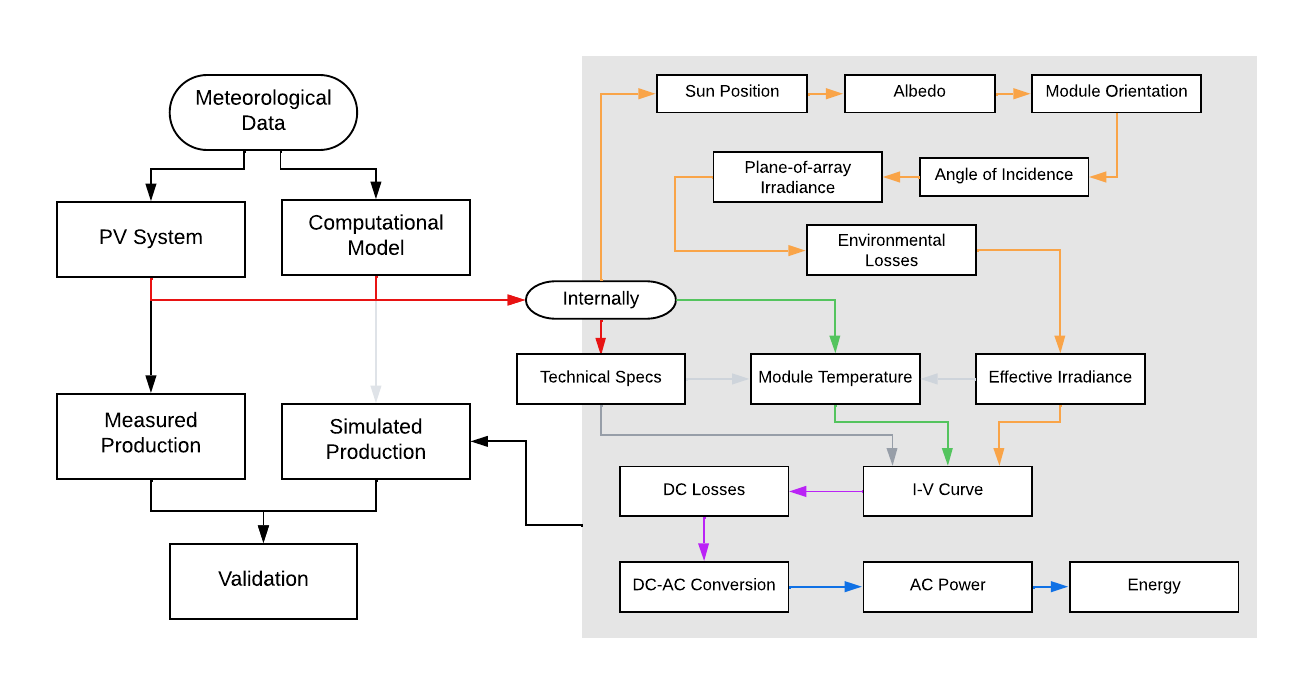}
\caption{Modeling of a PV system according to PVPMC \cite{ref:C2R12, ref:C2R20}. The grey arrows indicates the technical and electrical specifications; in yellow the effective irradiance modeling; in green the estimation of the PV panel temperature; in purple the DC production; and in blue the AC production.}
\label{fig:C2F1}
\end{figure}

The construction of the computational model follows the standard PV modeling steps recommended by PVPMC (see Figure \ref{fig:C2F1}).

The process begins with organizing the data to be used. The information extracted from the Meteocontrol monitoring system includes meteorological and production values from August 1, 2019, to February 28, 2021, with a five-minute resolution. The data is preprocessed to eliminate missing data concerning time. Each parameter has 166 464 data points. The parameters are detailed for System A and System B (recall Table \ref{tab:C1T5}) and are: global horizontal irradiance (GHI) from a pyranometer, plane-of-array (POA) irradiance from areference cell, ambient temperature, PV panel temperature, DC and AC voltage, DC and AC current, DC and AC power, and daily accumulated energy.

The chosen irradiance measurement is the one reported by the reference cell, as this instrument has the same material as the PV panels (i.e., monocrystalline silicon) and the same inclination and azimuthal angle (i.e., 10 and 180$^{\circ}$, respectively), which corresponds to the effective irradiance for the PV effect. In contrast, the pyranometer captures the GHI, and decomposition and transposition models must be used to calculate the POA irradiance.

Irradiance data $\leq$ 1.5 W/m$^2$ is set to zero; although these data have low magnitude, they can cause a significant deviation in the simulated values (i.e., butterfly effect). Typically, these data are located around sunrise and sunset.

In addition to the meteorological parameters already mentioned, the geographic variables of longitude, latitude, time zone, and altitude are defined (4.6024, -74.0674, America/Bogotá and 2 624 m, respectively).

The computational design of the PV system begins with the definition of the inverters. PVlib has databases of commercial inverters that include the main technical and operational parameters. The technical specifications of the inverter of the studied PV system is obtained from the California Energy Comission (CEC) inverter repository.

The definition of the PV panels was done manually to include parameters that were not available in the PVlib database. Table \ref{tab:C2T2} presents the PV panel parameters required for subsequent calculations.

To finalize the design of the PV system, the strings for each inverter are defined. The required parameters are the tilt, azimuth, PV panels connected in series, and strings connected in parallel per inverter. These data was provided in Table \ref{tab:C1T5}.

\begin{table}[]
\centering
\caption{Technical parameters of the PV panel LG Electronics LG400N2W-A5.}
\label{tab:C2T2}
\begin{tabular}{cccl}
\hline
Parameter & Value & Units & Description \\ \hline
$\alpha _{sc}$ & 3.14 $\cdot 10^{-3}$ & A/$^\circ$C & Temperature coefficient of the short-circuit current \\
$a_{ref}$ & 1.82 & V & \begin{tabular}[c]{@{}l@{}}Product of the diode ideality factor, the number of cells\\ in series, and the thermal voltage of the cell under STC\end{tabular} \\
$I_{L_{ref}}$ & 10.48 & A & Current generated by light under STC \\
$I_{o_{ref}}$ & 1.80 $\cdot 10^{-11}$ & A & Diode reverse saturation current under STC \\
$R_{{sh}_{ref}}$ & 293.80 & $\Omega$ & Shunt resistance under STC \\
$R_s$ & 0.31 & $\Omega$ & Series resistance under STC \\
Adjust & 9.38 & \% & Adjustment of $\alpha_{sc}$ \\
EgRef & 1.12 & eV & Bandgap energy at STC temperature \\
dEgdT & -2.67 $\cdot 10^-4$ & 1/K & Temperature dependence of EgRef under STC \\ \hline
\end{tabular}
\end{table}

The modeling of DC production continues by estimating the values of the five parameters of the single-diode equivalent circuit proposed by \cite{ref:C2R4}. However, the function developed by the CEC includes the $Adjust$ parameter in the estimation of these five parameters.

The equivalent circuit estimates the electrical behavior of a PV panel as follows:

\begin{equation}
\label{eq:C2E1}
I = I_L -I_o \left( \text{exp} \left(\frac{V + IR_s}{a} \right) -1 \right) - \frac{V + IR_s}{R_{sh}}
\end{equation}

Being $I_L$ the photoluminescence current, $I_o$ the reverse saturation current causing diffusion phenomena in the P-N junction, $a$ the modified ideality factor, and the resistances representing the non-ideal behavior of the panel: shunt resistance $R_{sh}$ limiting current flow caused by manufacturing defects, and series resistance $R_s$ reducing efficiency due to heat generation \cite{ref:C2R4, ref:C2R21}.

The difference between the CEC five-parameter model and the one proposed in \cite{ref:C2R4} lies in the dependency of $I_L$ on the temperature coefficient of $I_{sc}$, denoted as $\alpha_{sc}$ (see Eq. \ref{eq:C2E2}, where $G$ is the irradiance).

\begin{equation}
\label{eq:C2E2}
    I_L = \frac{G}{G_{STC}} \left( I_{sc,STC} + \alpha_{sc} \left( T_{cell} - T_{cell,STC} \right) \right)
\end{equation}

The CEC model adjusts the temperature coefficient $\alpha_{sc}$ with the parameter $Adjust$ so that the measurement and estimation of the MPP temperature coefficient match. According to \cite{ref:C2R21}, adjusting $\alpha_{sc}$ (i.e., $\alpha'_{sc}$) does not significantly affect the shape of the I-V curve, especially near the MPP. However, it is important because in simulations, PV panels operate near the MPP, so it is crucial to ensure that the MPP temperature dependence matches the value provided by the manufacturer \cite{ref:C2R21}.

\begin{equation}
\label{eq:C2E3}
\alpha'_{sc} = \alpha_{sc} \left( 1 - \frac{Adjust}{100} \right)  \;
\end{equation}

With the five parameters obtained, the single-diode model is used to solve the equivalent circuit equation (Eq. \ref{eq:C2E1}) and obtain the corresponding I-V curve. In addition to the five parameters, it is necessary to define the method for calculating the points on the curve. For computational efficiency, the chosen method to solve the implicit equation of the single-diode model is the Lambert-W function \cite{ref:C2R22}. The single-diode model estimates the DC production at MPP, i.e., $I_{DC}, V_{DC},$ and $P_{DC}$. The results are for a single PV panel. Therefore, these values must be scaled according to the array of panels for each inverter.

To estimate the AC production the SNL model (see Eq. \ref{eq:C2E4}--\ref{eq:C2E7}) considering the technical specifications of the inverter, $V_{DC}$ and $P_{DC}$ \cite{ref:C2R23}.

\begin{equation}
\label{eq:C2E4}
P_{AC} = \left( \frac{P_{AC_0}}{A-B} - C \left(A-B \right) \right) \cdot \left( P_{DC} - B \right) - C \left( P_{DC} - B \right)^2  \;
\end{equation}

\begin{equation}
\label{eq:C2E5}
A = P_{DC_0} \left( 1 + C_1 \left(V_{DC} - V_{DC_0} \right) \right)  \;
\end{equation}

\begin{equation}
\label{eq:C2E6}
B = P_{S_0} \left( 1 + C_2 \left(V_{DC} - V_{DC_0} \right) \right)  \;
\end{equation}

\begin{equation}
\label{eq:C2E7}
C = C_0 \left( 1 + C_3 \left(V_{DC} - V_{DC_0} \right) \right)  \;
\end{equation}

\begin{table}[]
\centering
\caption{Parameters for estimating AC power with SNL model). Adapted from \cite{ref:C2R23}.}
\label{tab:C2T3}
\begin{tabular}{ccl}
\hline
Parameter & Units & Description \\ \hline
$P_{AC}$ & W & \begin{tabular}[c]{@{}l@{}}AC power output from the inverter based on DC power and voltage input\end{tabular} \\
$P_{DC}$ & W & DC power input to the inverter \\
$V_{DC}$ & V & DC voltage input to the inverter \\
$P_{AC_0}$ & W & \begin{tabular}[c]{@{}l@{}}Maximum AC power output of the inverter under nominal operating\\condition; assumed as the upper saturation limit\end{tabular} \\
$P_{DC_0}$ & W & \begin{tabular}[c]{@{}l@{}}DC power at which nominal AC power is achieved under nominal\\ operating condition\end{tabular} \\
$V_{DC_0}$ & V & \begin{tabular}[c]{@{}l@{}}DC voltage at which nominal AC power is achieved under nominal\\operating condition\end{tabular} \\
$P_{S_0}$ & W & \begin{tabular}[c]{@{}l@{}}DC power required to start inverter operation; influences inverter\\efficiency at low power levels\end{tabular} \\
$C_0$ & 1/W & \begin{tabular}[c]{@{}l@{}}Defines the curvature of the DC-AC power relationship under nominal\\operating condition\end{tabular} \\
$C_1$, $C_2$, $C_3$ & 1/V & \begin{tabular}[c]{@{}l@{}}Empirical coefficients relating linear variation of $P_{DC_0}$ with $V_{DC}$\end{tabular} \\ \hline
\end{tabular}
\end{table}

According to \cite{ref:C2R23}, the main advantage of the SNL model is its consideration of sources causing non-linearity between DC and AC power for a given DC voltage. This results in a variable efficiency of the inverter, which is more accurate than assuming a linear efficiency. Some of the loss sources affecting inverter efficiency that the SNL model takes into account include: inverter self-consumption through the parameter $P_{S_0}$, losses proportional to $P_{AC}$ due to fixed voltage drops in semiconductors, and ohmic losses due to wiring

The accuracy of the SNL model depends on the available data to determine performance parameters; with the use of all required parameters, the model has an approximate error of 0.1\% between modeled and measured inverter efficiency \cite{ref:C2R23}. \cite{ref:C2R24} also validate the SNL model against other mathematical models for estimating AC power.

\subsection{Loss Estimation}
\label{subsec:C2S2S3}
The losses to be estimated are: soiling, degradation, DC and AC wiring, and inverter efficiency. The strategies used to estimate these loss factors differ from those employed by PVlib. PVlib and PVWatts utilize the model developed by NREL, where losses are not explicitly modeled but provided as a percentage. These percentages are based on assumptions and historical studies \cite{ref:C2R25}. Lastly, each loss factor $L_i$ contributes to determining the total loss value $L_{total}$ that derates the energy.

\begin{equation}
\label{eq:C2E8}
L_{total} = 100 \left( 1 - \prod_{i} \left( 1 - \frac{L_i}{100} \right) \right)  \;
\end{equation}

The strategy proposed by NREL is suitable for commercial models since typical values of loss factors are considered to enhance energy estimation accuracy. However, this method is not ideal for a research approach as losses are not dynamically associated with production. In contrast, the project strategy employs algorithms to estimate loss factor values based on monitored production data, ensuring dynamic adjustment of losses according to meteorological and operational conditions of the PV system.

\subsubsection{Soiling}
\label{subsec:C2S2S3S1}
The algorithm to estimate soiling losses is based on the method proposed by \cite{ref:C2R26, ref:C2R27, ref:C2R28}. The method starts by defining a performance metric $PM$ that captures all parameters affecting the electrical output of the PV system. \cite{ref:C2R26} proposes a $PM$ corrected for temperature and normalized with respect to insolation. Temperature and insolation are the parameters that most affect production; thus, correcting and normalizing with respect to these parameters allows distinguishing disturbances from other sources.

\begin{equation}
\label{eq:C2E9}
PM = \frac{E}{H}  \;
\end{equation}

Being $H$ the insolation and $E$ the energy, both in daily resolution. $H$ is obtained by integrating the irradiance for each day, while $E$ is calculated from the temperature-corrected AC power $T_{AC,TC}$. The mathematical model for temperature correction $T_{corr}$ is defined and validated in \cite{ref:C2R29, ref:C2R30}.

\begin{equation}
\label{eq:C2E11}
P_{AC,TC} = \frac{P_{AC}}{T_{corr}} \;
\end{equation}

\begin{equation}
\label{eq:C2E10}
T_{corr} = 1 + \alpha_{sc} \left( T_{cell} - T_{cell,STC} \right)  \;
\end{equation}

The data used to estimate the $PM$ indicator is monitored during active production hours, i.e., from 6:00 to 18:00. Unlike \cite{ref:C2R26, ref:C2R27}, who only consider irradiance values $\geq$ 500 W/m$^2$, the algorithm is adapted to use all POA irradiance. The advantage of the daily solution is that it eliminates intraday variations of $PM$ while adequately capturing soiling patterns \cite{ref:C2R27}.

\begin{figure}[]
\centering
\includegraphics[width=0.90\textwidth]{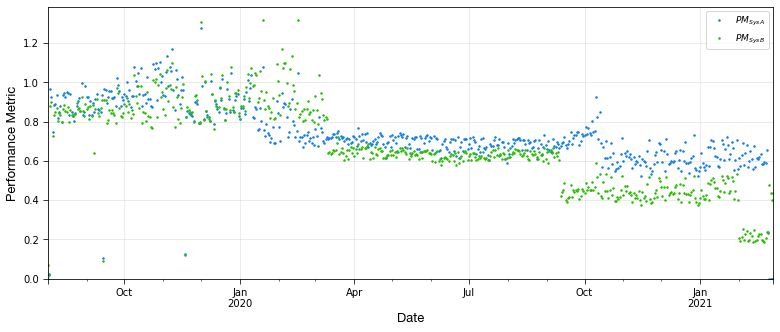}
\caption{Behavior of the defined performance index $PM_{norm}$. Blue represents System A and green represents System B. The color convention is used throughout the paper.}
\label{fig:C2F3}
\end{figure}

Finally, the $PM$ is normalized relative to the 95th percentile of these same values (Eq. \ref{eq:C2E9}) to obtain a dimensionless metric $PM_{norm}$, minimizing the impact of biases in the model and potential outliers at the extremes (see Figure \ref{fig:C2F3}).

The algorithm proceeds with automatic detection of cleaning events without the need for rain data or cleaning records, dividing the dataset into intervals or soiling periods \cite{ref:C2R26, ref:C2R27}. To achieve this, the patterns of the centered 14-day moving median (a filter that reduces noise influence and preserves step changes in the data) of $PM_{norm}$ are first analyzed. Subsequently, differences $\Delta$ between neighbor values.

\begin{equation}
\label{eq:C2E12}
\Delta = PM_{norm,i} - PM_{norm,i-1}\;
\end{equation}

Finally, the identification of cleaning events occurs when the absolute value of the difference series $|\Delta|$ is greater than $Q_3 + 1.5 IQR$, where $Q_3$ is the third quartile (75th percentile) and $IQR$ is the interquartile range. By definition, a value in a statistical distribution is classified as an outlier if it exceeds this mathematical relationship; in other words, a significant positive change between neighboring data points of $PM_{norm}$ indicates improved performance and thus a cleaning event.

\begin{figure}[]
\centering
\begin{subfigure}{1\textwidth}
   \centering
   \includegraphics[width=0.90\linewidth]{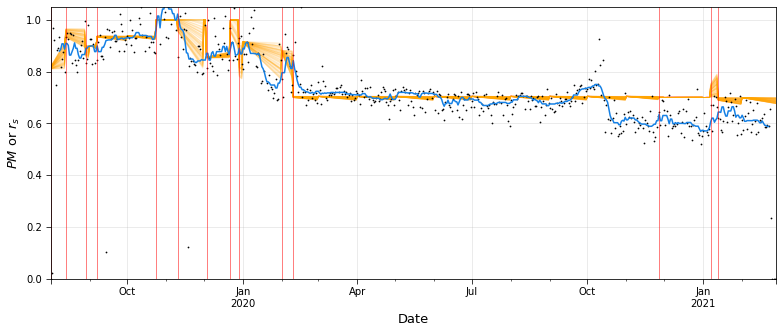}
   \label{fig:C2F5a} 
\end{subfigure}

\begin{subfigure}{1\textwidth}
   \centering
   \includegraphics[width=0.90\linewidth]{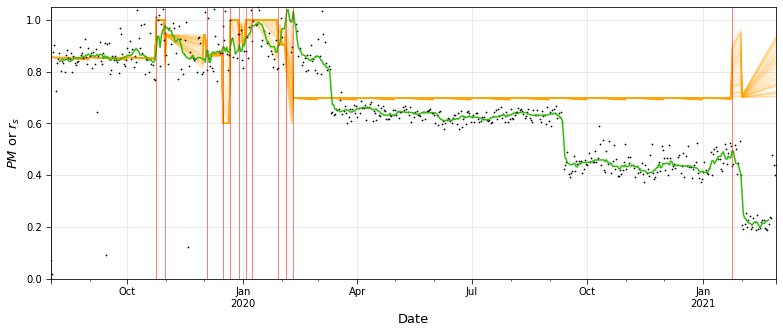}
   \label{fig:C2F5b} 
\end{subfigure}
\caption{Detection of cleaning events for System A (top) and System B (bottom). Each red line represents a cleaning event; the period between events is referred to as soiling interval. In orange, the Monte Carlo simulation for stochastic generation of possible soiling profiles.}
\label{fig:C2F5}
\end{figure}

The next step involves determining the slope of each soiling interval and the cleaning magnitude between them (see Figure \ref{fig:C2F5}). The slope is determined using the Theil-Sen estimator, which robustly fits the slope using the median of all slopes between pairs of points \cite{ref:C2R27}. The Theil-Sen estimator also calculates the 95\% confidence interval for the slope. On the other hand, the magnitude of each cleaning event is calculated based on the difference between the fitting value at the beginning of an interval and the final fitting value of the previous interval \cite{ref:C2R27}. 

The algorithm proposed by \cite{ref:C2R26, ref:C2R27} continues with the Monte Carlo simulation to stochastically generate soiling profiles and obtain a more reliable approximation of the soiling rate $PM_{norm}$ (see Figure \ref{fig:C2F5}). The Monte Carlo simulation involves randomly selecting the slope of each soiling interval based on the 95\% confidence interval calculated using the Theil-Sen estimator. 

Finally, soiling losses weighted by insolation $r_{s,H}$ are estimated. Once the Monte Carlo simulation is performed and the daily values of $PM_{norm}$ are obtained, $r_{s,H}$ is determined to normalize the performance metric $PM_{norm}$ with respect to daily levels of insolation $H_{i}$. \cite{ref:C2R27} defines $r_{s,H}$ as follows:

\begin{equation}
\label{eq:C2E13}
r_{s,H} = \frac{\sum_{i} H_{i} \cdot P_{norm,i}}{\sum_{i} H_{i}} \;
\end{equation}

According to \cite{ref:C2R27}, the benefit of estimating $r_{s,H}$ is that the soiling-to-insolation weighted ratio captures the expected energy loss due to the soiling whether during periods of high or low insolation. This aspect is not captured by simply averaging the soiling ratio or estimating the soiling rate as a percentage of loss.

\subsubsection{Degradation}
\label{subsec:C2S2S3S2}
\cite{ref:C2R31} studies the main methodologies to estimate PV degradation by creating synthetic datasets with unit degradation and applying three different methods that consider the amount of annual data available, noise, and seasonal influences. These are:

\begin{itemize}
    \item \textbf{Standard least squares}. This approach uses all data points in a single regression, minimizing the difference between the model and measured data.
    
    \item \textbf{Robust regression}. It seeks to minimize the Huber loss function, which increases quadratically for small errors and linearly for large errors, mitigating the influence of outliers.
    
    \item \textbf{Year-over-year (YOY)}. \cite{ref:C2R31, ref:C2R32, ref:C2R33} propose determining degradation by leveraging modeled clear-sky irradiance data instead of in-situ irradiance sensor data. A line is fitted between two points (hours, days, weeks, or months) in subsequent years to determine a degradation rate for these specific points. The result is a distribution of degradation rates where the central tendency is considered representative of the long-term performance behavior.
\end{itemize}

The strengths of regression approaches include a lower 95\% confidence interval for a smaller amount of annual data available, less noise, and seasonality. On the other hand, the advantage of the YOY approach is its lower sensitivity to outliers, seasonality, soiling, and leverage of the initial and final points of analysis \cite{ref:C2R31, ref:C2R32, ref:C2R33}. However, YOY requires a minimum of two years of data, and this requirement is not met for this research. Therefore, the strategy is to estimate the annual degradation rate based on the median of ten annual degradation rate values reported in the literature. Then, the estimated annual degradation rate is linearly distributed over the time window considered. The estimated degradation rate is -0.5\% per year, which aligns with typical degradation reported in the literature.

\subsubsection{DC and AC Wiring}
\label{subsec:C2S2S3S3}
The internal resistance of wiring in PV systems results in ohmic losses due to voltage drop during current flow. These losses increase with temperature and are typically estimated to be $\leq$ 3\% per year \cite{ref:C2R25, ref:C2R34}.

Ohmic losses in DC wiring occur in the wiring between PV panels and the input terminals of the inverter. On the other hand, AC wiring includes all wires downstream from the inverter to the connection point. Ohmic losses in AC wiring are typically lower than those in DC wiring and are further reduced in PV systems with central inverters \cite{ref:C2R25}.

\begin{equation}
\label{eq:C2E15}
\text{Ohmic Losses} = I^2 \cdot \rho \frac{\ell}{A_T}
\end{equation}

Where $I$ is the current, $\rho$ is the electrical resistivity (0.0171 $\Omega$ mm$^2$/m for copper), $\ell$ is the wire length, and $A_T$ is the cross-sectional area of the cable.

\subsubsection{Inverter Performance}
\label{subsec:C2S2S3S4}
The estimation of inverter losses is obtained by relating the DC and AC power of the system. The DC power is obtained from the Meteocontrol monitoring system, and the AC power is calculated using the SNL model (recall Eq. \ref{eq:C2E4}) \cite{ref:C2R23, ref:C2R24}.

\begin{equation}
\label{eq:C2E16}
\text{Inverter Losses} = \left( 1 - \frac{P_{AC}}{P_{DC}} \right) \cdot 100 \;
\end{equation}

\subsection{Incorporation of Losses into Modeling}
\label{subsec:C2S2S4}
The losses estimated in Section \ref{subsec:C2S2S3} are incorporated into production according to their occurrence in the logical operation process of the PV system (see Figure \ref{fig:C3F6}).

\section{Implementation}
\label{sec-5}

\subsection{Synthetic Dataset}
\label{subsec:C3S2S1}
The main advantage of a synthetic dataset is that it allows understanding the state of the PV system by defining each data point as normal or anomalous. The methodology proposed in \cite{SalazarPena2023} is used to generate a synthetic dataset of POA irradiance and ambient temperature. The synthetic data is subject to the actual production of the PV system and the weather conditions of a specific date; it reinforces the consistency of the irradiance fluctuation dynamics with the meteorological state. In addition, a five-minute resolution allows understanding of intra-day behaviors. The synthetic generation of ambient temperature $T_{amb}$ and PV cell temperature $T_{cell}$ is based on the $\pm$2.5\% range of synthetic irradiance for each data point. Each synthetic temperature value is derived from the Gaussian distribution of temperatures within this range and considering the climatic categorization.

\subsection{Climatic Categorization}
\label{subsec:C3S2S1S1}
The climatic state and its dynamic nature are characterized using the clearness index $k$ defined by \cite{ref:C3R14} as the ratio between the POA irradiance $G_{POA}$ and the irradiance under clear sky conditions $G_{cs}$. The values of $G_{cs}$ are obtained using the PVlib library. Using this index, \cite{ref:C3R13} successfully estimates the meteorological conditions of the day.

\begin{table}[]
{\centering
\caption{Climate condition according to the value of $k$. Adapted from \cite{ref:C3R13}.}
\label{tab:C3T1}
\resizebox{\textwidth}{!}{%
\begin{tabular}{cccl}
\hline
Range of $k$ & Denomination$^a$ & Climate Condition & Description \\ \hline
[0 – 0.2] & SC1 & Completely overcast & Thick clouds with little to no fluctuation \\
(0.2 – 0.4] & SC2 & Mostly overcast & Alternating clouds and clear sky, some or large fluctuations \\
(0.4 – 0.6] & SC3 & Partly cloudy & Thin clouds with some or large fluctuations and/or haze \\
(0.6 – 0.67] & SC4 & Mostly clear & No clouds but some haze \\
(0.67 – 1] & SC5 & Completely clear & Clear sky \\ \hline
\end{tabular}}
}
\newline $^a$ \emph{Sky condition} (SC).
\end{table}

\begin{figure}[]
\centering
\includegraphics[width=0.90\textwidth]{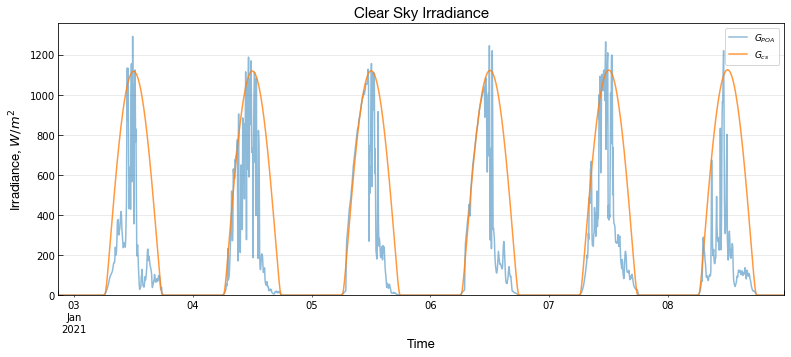}
\caption{Comparison between $G_{POA}$ in blue and $G_{cs}$ in orange.}
\label{fig:C3F1}
\end{figure}

Figure \ref{fig:C3F1} shows the behavior of these irradiances for the first week of January 2021. The $G_{cs}$ values are the theoretically maximum. However, achieving a profile with little variability is uncommon. Instances where $G_{POA}$ exceeds $G_{cs}$ may be due to reflections in the reference cell.

\begin{figure}[]
\centering
\includegraphics[width=0.90\textwidth]{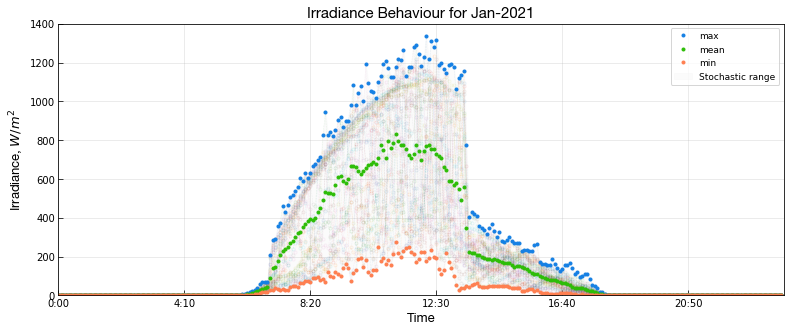}
\caption{Maximum, average, and minimum behavior of daily irradiance for January 2021.}
\label{fig:C3F2}
\end{figure}

In Figure \ref{fig:C3F2}, daily $G_{POA}$ measurements throughout January 2021 are presented, including statistics for maximum, mean, and minimum. The gray area indicates the range within which proposed methodology in \cite{SalazarPena2023} can generate synthetic $G_{POA}$ values while maintaining climatic physics.

To improve accuracy, $k$ is used to categorize the weather behavior of each day of each month.. This approach allows the algorithm to better capture the meteorological dynamics of each set of days and ensures that the generated synthetic value is physically consistent. Figure \ref{fig:C3F2} displays the irradiance measurements for each dataset categorized according to $k$. For January 2021, no days were classified as SC1 (see Table \ref{tab:C3T1}). As the climatic conditions approach a clear-sky condition, the irradiance values become closer to each other.

\begin{figure}[]
\centering
\begin{subfigure}{0.475\textwidth}
   \centering
   \includegraphics[width=\linewidth]{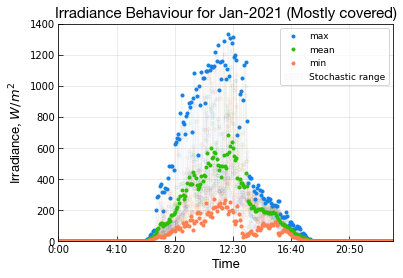}
   \label{fig:C3F3a} 
\end{subfigure}
\begin{subfigure}{0.475\textwidth}
   \centering
   \includegraphics[width=\linewidth]{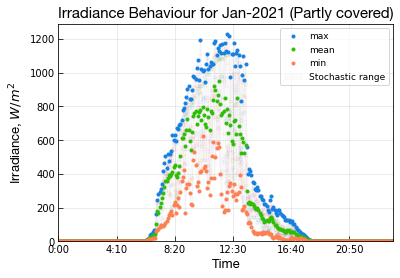}
   \label{fig:C3F3b} 
\end{subfigure}

\begin{subfigure}{0.475\textwidth}
   \centering
   \includegraphics[width=\linewidth]{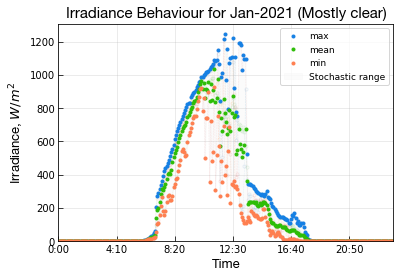}
   \label{fig:C3F3c} 
\end{subfigure}
\begin{subfigure}{0.475\textwidth}
   \centering
   \includegraphics[width=\linewidth]{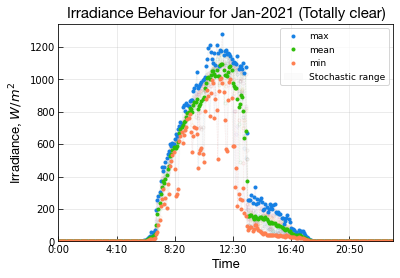}
   \label{fig:C3F3d} 
\end{subfigure}
\caption{Daily irradiance for January 2021 categorized according to the sky condition given by $k$.}
\label{fig:C3F3}
\end{figure}

\subsection{Fault Generation}
\label{subsec:C3S2S2}
According to \cite{ref:C3R1, ref:C3R2, ref:C3R17}, faults are categorized as:

\begin{itemize}
    \item \textbf{Permanent:} Faults that affect the integrity of components; they are resolved by replacing the defective equipment (e.g., line-to-line, arc fault, bypass diode, delamination, microcracks).
    
    \item \textbf{Temporary:} Faults that can be trivially removed using a specific method or tool (e.g., shading, soiling, pollution, hot spot).
    
    \item \textbf{Incipient:} Temporary faults that, if not properly addressed, can lead to permanent faults (e.g., degradation, corrosion, interconnection, partial damage).
\end{itemize}

A recognition of potential failures affecting the PV system under study is conducted. As a result, the failures to be included in the synthetic datasets are presented in Table \ref{tab:C3T3}.

\begin{table}[]
{\centering
\caption{Description of faults to consider in the generation of synthetic data for the PV system at Universidad de los Andes.}
\label{tab:C3T3}
\resizebox{\textwidth}{!}{%
\begin{tabular}{clccc}
\hline
Fault & Description & Range (\%) & Affects & Ref \\ \hline
Soft Shading$^a$ (SS) & \begin{tabular}[c]{@{}l@{}}Movement of clouds and shadows from trees\end{tabular} & 10--20 & $P_{DC}$ & \cite{ref:C3R18} \\
Hard Shading $^a$ (HS) & \begin{tabular}[c]{@{}l@{}}Shadow of a solid material such as buildings, bird droppings, or soiling\end{tabular} & 10--70 & $P_{DC}$ & \cite{ref:C3R17} \\
Soiling $^a$ (SOI) & \begin{tabular}[c]{@{}l@{}}Resulting from accumulation of snow, soiling, or any other particles\end{tabular} & 8--12 & $P_{DC}$ & \cite{ref:C3R18} \\
Hot Spot (HSp) & \begin{tabular}[c]{@{}l@{}}Regions with elevated cell temperature; occurs when operating in\\reverse bias (i.e., instead of producing energy, they dissipate it)\end{tabular} & 2$^b$ & $T_{cell}$, $P_{DC}$ & \cite{ref:C3R19} \\
Degradation (DEG) & \begin{tabular}[c]{@{}l@{}}Deterioration of PV cell integrity (e.g., delamination, bubbles,\\anti-reflection coating deficit)\end{tabular} & 0--50 & $P_{DC}$ & \cite{ref:C3R17} \\
Cell Cracks (CC) & \begin{tabular}[c]{@{}l@{}}Result from stresses during manufacturing, mishandling during\\ transport and installation, or environmental factors\end{tabular} & 0.9--42.8 & $T_{cell}$, $P_{DC}$ & \cite{ref:C3R18} \\
Line-line (LL) & \begin{tabular}[c]{@{}l@{}}Accidental short circuit between two points of different\\potential in an array\end{tabular} & 20--80 & $V_{oc}$ & \cite{ref:C3R2} \\
Line-ground (GF) & \begin{tabular}[c]{@{}l@{}}Accidental short circuit between one or more\\ current-carrying conductors and ground\end{tabular} & 0--51.9 & $I_{sc}$, $P_{DC}$ & \cite{ref:C3R2} \\
Series Arc Fault (SAF) & \begin{tabular}[c]{@{}l@{}}Discontinuity in current-carrying conductors due to cell damage,\\connector corrosion, or solder joint disconnections\end{tabular} & 0--62.5$^c$, 0--80$^d$ & $I_{DC}$, $V_{DC}$ & \cite{ref:C3R2} \\
Parallel Arc Fault (PAF) & \begin{tabular}[c]{@{}l@{}}Arc between two parallel conductors due to insulation breakdown\\or as a secondary effect from series arc\end{tabular} & 0--87.5 & $V_{DC}$ & \cite{ref:C3R2} \\
Inverter (IF) & \begin{tabular}[c]{@{}l@{}}Overvoltage and harmonic distortion of voltage due to thermomechanical\\ stress on power switches and failure of capacitors and control circuit\end{tabular} & 0--100 & $P_{AC}$ & \cite{ref:C3R20} \\ \hline
\end{tabular}}
}
\newline $^a$ Bypass diodes mitigate these effects and improve energy flow, but introduce peaks in I-V curves that hinder reaching the global peak of the MPPT \cite{ref:C3R1, ref:C3R2, ref:C3R17}.
$^b$ Losses of 2\% per string according to \cite{ref:C2R28}.
$^c$ Affects $I_{DC}$.
$^d$ Affects $V_{DC}$.
\end{table}

The fault values are randomly generated using a bounded uniform distribution according to the percentage impact range of each fault listed in Table \ref{tab:C3T3}. Similarly, the number of data points to be affected is obtained using a uniform distribution, limited to 50\% of the total data to avoid class imbalance. The goal is to include known faults at specific time instances to represent the stochastic nature of fault occurrences in the PV system operation. Algorithm \ref{alg:C3A5} details the procedure followed for generating faults in the synthetic databases.

\begin{algorithm}[]
\SetAlgoLined
\DontPrintSemicolon
  \textbf{Input} Fault name (\emph{fault\_name}), min value (\emph{fault\_min}), max value (\emph{fault\_max}), synthetic data (\emph{synt})\newline
  \textbf{Output:} Fault value (\emph{fault\_val})

  \For{daily dataset in synt}{
    Define the percentage of data points to impact (\emph{percentage}) \tcp*{uniform(0, 0.5, 1)}
    Define which data points will be impacted (\emph{fault\_points}) \tcp*{uniform(0, len(synt), percentage*len(synt))}
  
    \For{index in synt}{
      Check if \emph{index} is in \emph{fault\_points} list\newline
      Define the fault impact value (\emph{fault\_imp}) \tcp*{uniform(fault\_min, fault\_max, 1)}
      Divide \emph{fault\_imp} by 100 to get the percentage
    }
    Add a new \emph{fault\_name} column in the \emph{synt} dataset to store the fault values
  }
\caption{Stochastic Fault Generation}
\label{alg:C3A5}
\end{algorithm}

\subsection{Database Construction}
\label{subsec:C3S2S3}
The database construction starts with the process of generating synthetic $G_{POA}$, $T_{amb}$ and $T_{cell}$. To have a representative dataset, a total of 2 039 040 data points are generated with five-minute resolution (i.e., approximately 19 years) of synthetic $G_{POA}$ for System A and System B.

Based on the quantified losses of the PV system under study, Algorithm \ref{alg:C3A5} is employed to define a loss factor value. This algorithm consists of four steps:

\begin{enumerate}
    \item Take the number of days according to the month of synthetic the dataset (e.g., 31 days for Jan-2021).
    
    \item Select a random day within the range given by month and year (e.g., Jan 10, 2021).
    
    \item Locate the randomly selected day in the loss data and take the respective value for each loss factor.
    
    \item The taken loss values are stored in a separate list for each factor, and the value is repeated for the number of days according to the month of synthetic data.
    
    \item Store the augmented lists in the synthetic database.
\end{enumerate}

\begin{figure}[]
    \centering
    \includegraphics[width=0.60\textwidth]{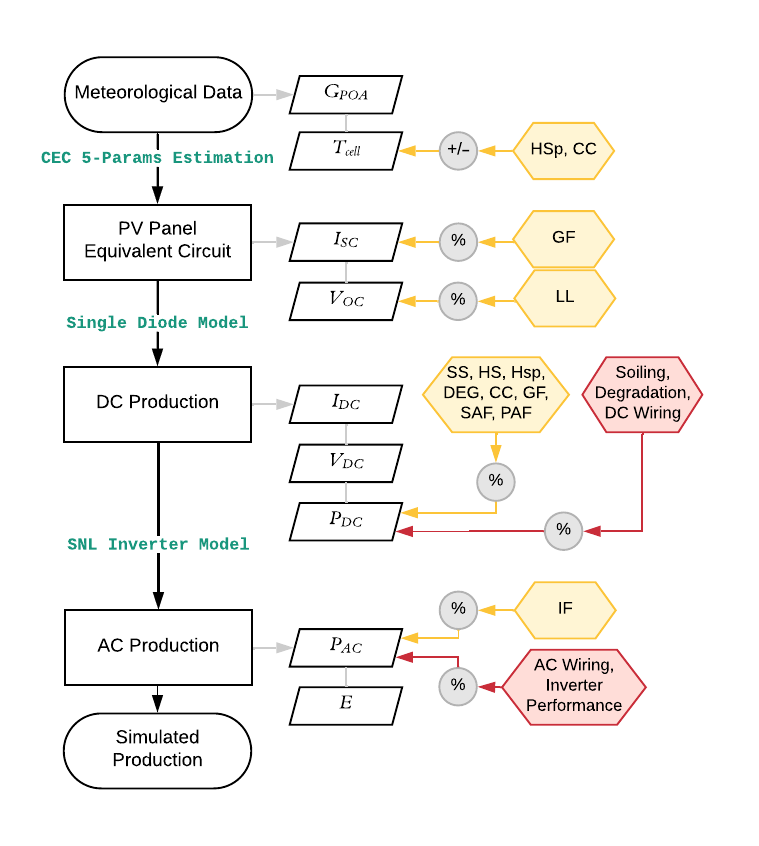}
    \caption{Diagram of the incorporation of losses and faults in synthetic production according to the logical operation process of the PV system. Faults are shown in yellow and losses in red.}
    \label{fig:C3F6}
\end{figure}

The objective is to include known faults at specific time instances to represent the stochastic nature of fault occurrence in PV system operation. Figure \ref{fig:C3F6} illustrates the procedure for adding faults to the synthetic dataset.

\subsection{AI Agent}
\label{subsec:C3S2S4}
Fault detection is performed using a feedforward ANN designed with the PyTorch library and executed on a high-performance computing machine (recall Section \ref{subsec:C1S3S3}).

The ANN is configured with ten input parameters (i.e., $G_{POA}$, $T_{amb}$, $T_{cell}$, hour, month, $k$, $I_{DC}$, $V_{DC}$, $P_{DC}$, and $P_{AC}$) to capture representative information about the meteorological and operational conditions of the PV system. It has a single output parameter that provides information about the equipment's status (i.e., $I_L$, $I_{sc}$, $V_{oc}$, PV panel efficiency $\eta_{cell}$, and inverter efficiency $\eta_{inv}$) In other words, there is a specific ANN for each output parameter.

The inputs are normalized within the range [0--1] to standardize the variability across each set of information and improve the convergence speed, performance, and training stability.

The architecture of each ANN consists of two hidden layers: the first layer followed by the Sigmoid activation function and the second by LeakyReLU. Activation functions introduce nonlinearities into the data behavior.

In addition to the number of hidden layers and the established activation functions, the following hyperparameters are defined: 20\% dropout, 50 epochs, batch size of 5 000, mean squared error (MSE) as loss function, Adam optimizer, weight initialization using He method, and a learning rate (LR) of 0.1 with ReduceLROnPlateau scheduler. PyTorch's ReduceLROnPlateau scheduler dynamically adjusts LR based on validation metrics. In this case, LR is reduced by a factor of 0.1 if significant optimization is not detected in training over 5 epochs. LR resets to 0.01 if its value decreases to a magnitude of \( 10^{-7} \).

Furthermore, the ANN is trained with different configurations of the number of hidden layers as mentioned earlier, and the number of neurons per layer. For a two-hidden-layer architecture, four ANNs are designed with different node counts: 6, 9, 12, and 15 for both layers. These values are chosen due to their frequent use in the literature \cite{ref:C3R22, ref:C3R23, ref:C3R24, ref:C3R25}.

The results inferred by the ANN are compared against the target values, i.e., the synthetic data, to estimate the prediction error. Control metrics for the ANN during training and validation include root mean squared error (RMSE) and the \( R^2 \). Additionally, accuracy between the generated signal and measurement is evaluated using median absolute percentage error (MeAPE), which is a robust indicator resistant to outliers in comparison to the mean absolute percentage error (MAPE). The absolute percentage error (APE) is defined in Eq. \ref{eq:C2E17}, where $y$ is any parameter of analysis.

\begin{equation}
\label{eq:C2E17}
\text{APE} = \left| \frac{y_{\text{measured}} - y_{\text{modeled}}}{y_{\text{measured}}} \right| \cdot 100 \;
\end{equation}

Finally, a 5-fold cross-validation is conducted to assess the statistical analysis results, ensuring they are representative and independent of the training, validation, and test datasets. In the case of 5-fold cross-validation, the dataset is split into 80\% for training and 20\% for testing. Additionally, 85\% of the training data is used exclusively for training, while the remaining 15\% is employed for validation.

\subsection{Threshold Definition}
\label{subsec:C4S2S1}
The main drawback of conventional strategies for defining normal operation limits is the lack of consideration for dynamic meteorological nature and its influence on production, which can cause unexpected fluctuations.

The methodologies developed in Section \ref{subsec:C3S2S1} for generating synthetic irradiance do take into account climatic variability to ensure the physical consistency of the produced data. This procedure forms the basis of the alternatives that define the following thresholds:

\begin{itemize}
    \item $\mu \pm 3\sigma$. This alternative leverages the Gaussian or lognormal distribution of data \cite{SalazarPena2023}. For a Gaussian distribution, the interval $\pm 3\sigma$ encompasses 99.7\% of the data points. Even for distributions not following a Gaussian pattern, this interval still contains 88.9\% of the information \cite{ref:C3R6}.
    
    \item $Q_3 \pm 1.5 IQR$. This approach aims to define more reliable thresholds based on historical data while being less sensitive to outliers.
    
    \item Maximum and Minimum. This alternative verifies that the study signal falls within the historical maximum and minimum range.
\end{itemize}

Regardless of the strategy used, if the instantaneous data falls outside the corresponding normal operating range, it is classified as faulty. Otherwise, it is classified as normal operation. For simplicity, these are defined as 1 and 0, respectively.

\subsection{Evaluation Metrics for Accuracy}
\label{subsec:C4S2S2}

The confusion matrix is a tool used to evaluate the performance of a binary classification algorithm by indicating the quantity and percentage of correct and incorrect classifications. The confusion matrix is constructed based on the following parameters:

\begin{itemize}
    \item \textbf{True Positives (TP)}: The prediction correctly classified as positive according to the target value marked as positive, i.e., when both the prediction and the target indicate failure (1).
    
    \item \textbf{True Negatives (TN)}: The prediction correctly classified as negative according to the target value marked as negative, i.e., when both the prediction and the target indicate no failure (0).
    
    \item \textbf{False Positives (FP)}: The prediction incorrectly classified as positive according to the target value marked as negative, i.e., the prediction indicates 1 when the target is 0.

    \item \textbf{False Negatives (FN)}: The prediction incorrectly classified as negative according to the target value marked as positive, i.e., the prediction indicates 0 when the target is 1.
\end{itemize}

From these variables, the accuracy is defined as:

\begin{equation}
\label{eq:C4E1}
\text{accuracy} = \frac{TP + TN}{TP + FP + FN + TN} \;
\end{equation}

The accuracy indicates the proportion of correct classifications out of the total predictions made. This metric quantifies the accuracy of fault detection based on the thresholds defining the normal operational behavior of the PV system.

\section{Presentation and Analysis of Results}
\label{sec-6}

\subsection{Computational Modeling}
\label{sec:C2S3}

\subsubsection{Losses}
\label{subsec:C2S3S1}
\begin{figure}[]
\centering
\begin{subfigure}{1\textwidth}
   \centering
   \includegraphics[width=0.70\linewidth]{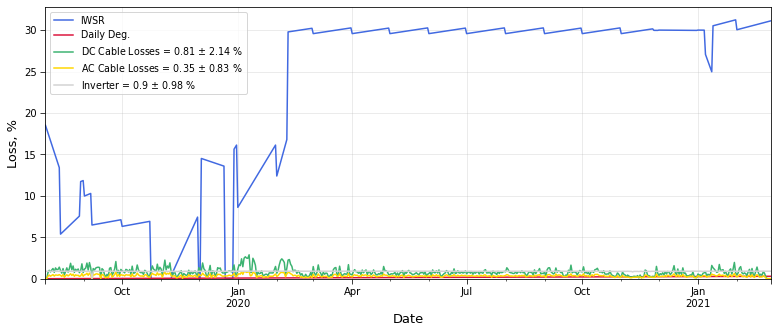}
   \label{fig:C2F7a} 
\end{subfigure}

\begin{subfigure}{1\textwidth}
   \centering
   \includegraphics[width=0.70\linewidth]{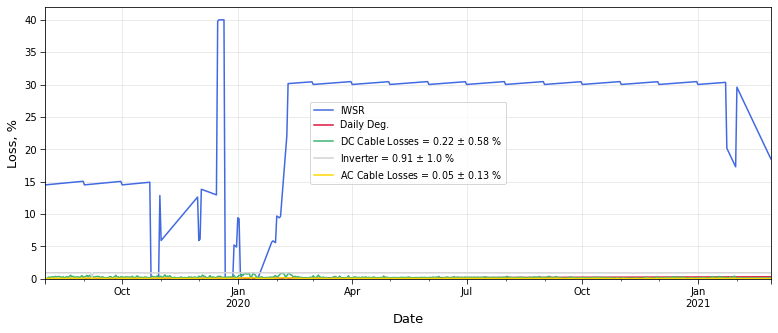}
   \label{fig:C2F7b} 
\end{subfigure}
\caption{Daily losses behavior of System A (top) and System B (bottom) of the PV system at Universidad de los Andes.}
\label{fig:C2F7}
\end{figure}

The estimation of daily losses according to the algorithms presented in Section \ref{subsec:C2S2S3} are shown in Figure \ref{fig:C2F7}. The blue curve represents soiling losses. Initially, losses of approximately 18\% and 20\% are estimated for System A and System B, respectively. Theoretically, these values are expected to be close to zero at the initial point, as this is when the PV system begins operation. One possible cause of this deviation is due to complications that the simulation faced in estimating losses as frequent missing data was verified through the Meteocontrol monitoring system in the first months of operation.

For System A and System B, it is observed that in January, soiling losses reach a minimum value of zero. This is consistent with the cleaning information provided in the preventive maintenance manual of the PV system under study, where periodic cleanings are scheduled for January and August. From this date onwards, it is observed that soiling losses gradually increase until reaching a maximum of approximately 30\%. This peak aligns with what is reported in the literature, where a short-term maximum of 30\% and a monthly maximum of approximately 20\% are established.

In the taxonomy of soiling loss profiles, significant positive slopes represent identified cleaning events, while negative slopes indicate the onset of soiling accumulation intervals; the slope magnitude indicates the particle density scale on the PV panels. Less significant positive slopes indicate mild cleanings due to rain and/or wind that lightly removes soiling particles; the slope magnitude indicates the intensity of rain and/or wind.

The green and yellow curves represent losses due to DC and AC wiring, respectively. The mean value for the evaluated time window is $0.8 \pm 2.1\%$ and $0.3 \pm 0.8\%$ for DC and AC cables of System A, respectively, and $0.2 \pm 0.5\%$ and $0.05 \pm 0.13\%$ for DC and AC cables of System B, respectively. These values are consistent with the literature, where mean losses of approximately 3\% are reported for DC cables and much lower values for AC cables. Additionally, AC wiring losses represent approximately 43\% and 23\% of those caused by DC cables in System A and System B, respectively. Finally, DC and AC wiring losses in System B are approximately 73\% and 86\% lower compared to System A, respectively; this is consistent with their installed capacity, demonstrating the dynamic nature of the estimated loss profile based on production data.

Degradation losses are represented by the red curve. As mentioned in Section \ref{subsec:C2S2S3S2}, the degradation rate is $0.5\%$ per year, resulting in degradation losses of approximately $0.8\%$ for the last day analyzed, i.e., February 28, 2021.

Finally, the gray curve indicates losses due to inverter performance, with an mean of $0.9 \pm 0.9\%$ for the time window studied across both systems. Factors associated with these losses include self-consumption of electrical resources by inverter components and ohmic losses. Given the crucial role of the inverter in a PV system, its efficiency is high (i.e., 98.4\% and 97.5\% for the inverters of System A and System B, respectively) resulting in effectively low magnitude losses due to inverter performance.

\subsubsection{DC Voltage}
\label{subsec:C2S3S2}
The statistical analysis shows a positive correlation between measurement and simulation. Regarding accuracy, the MeAPE for both systems is around 2\%, which fall within the ideal range (i.e., $\leq 5\%$).

\begin{figure}[]
\centering
\begin{subfigure}{1\textwidth}
   \centering
   \includegraphics[width=0.70\linewidth]{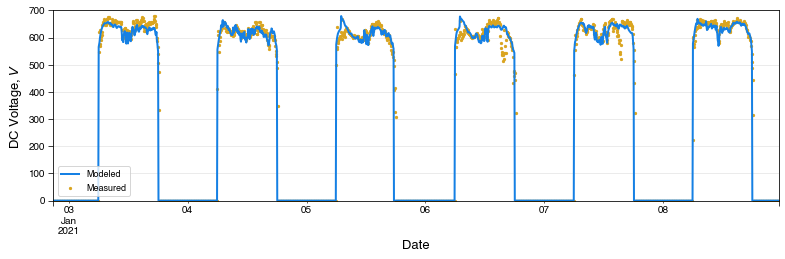}
   \label{fig:C2F8a} 
\end{subfigure}

\begin{subfigure}{1\textwidth}
   \centering
   \includegraphics[width=0.70\linewidth]{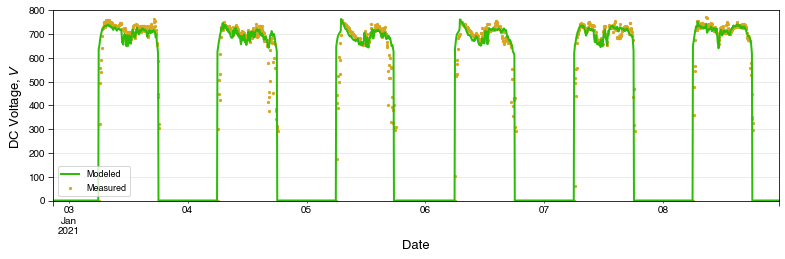}
   \label{fig:C2F8b} 
\end{subfigure}
\caption{Measured and modeled DC voltage for System A (top) and System B (bottom).}
\label{fig:C2F8}
\end{figure}

\subsubsection{DC Current}
\label{subsec:C2S3S3}
The $R^2$ reflects the positive relationship between measurement and the value estimated by the computational model. Additionally, RMSE indicates the simulation accuracy within the studied time window. However, MeAPE show a higher model inaccuracy compared to DC voltage, due to the erratic behavior of current and its high dependence on irradiance and temperature. The mean minimum APE for both systems is 0.003\%, confirming the high precision achieved by the computational model.

\begin{figure}[]
\centering
\begin{subfigure}{1\textwidth}
   \centering
   \includegraphics[width=0.70\linewidth]{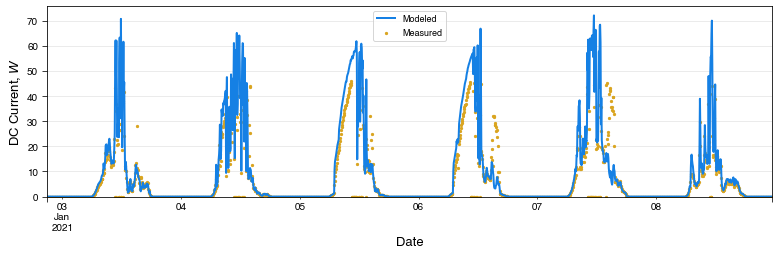}
   \label{fig:C2F9a} 
\end{subfigure}

\begin{subfigure}{1\textwidth}
   \centering
   \includegraphics[width=0.70\linewidth]{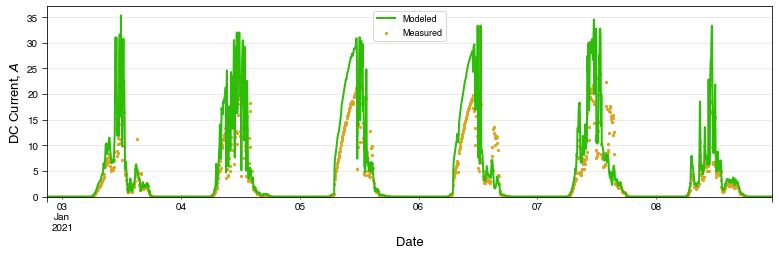}
   \label{fig:C2F9b} 
\end{subfigure}
\caption{Measured and modeled DC current for System A (top) and System B (bottom).}
\label{fig:C2F9}
\end{figure}

\subsubsection{DC and AC Power}
\label{subsec:C2S3S4}
The proportionality coefficient is positive and, being greater than 0.9, indicates a strong correlation between simulation and measurement. Similarly, RMSE demonstrates the accuracy between the values.

Both MAPE and MeAPE are very close to those obtained for DC current, which ratifies that the taxonomy of the simulated DC power profile is defined by the behavior of the DC current. This is also coherent as DC power encompasses the error propagated between DC voltage and DC current.

The minimum and maximum APE for System A are approximately 0.0003\% and 4.1\%, respectively. The magnitudes for System B is similar to that of System A.

\begin{figure}[]
\centering
\begin{subfigure}{1\textwidth}
   \centering
   \includegraphics[width=0.70\linewidth]{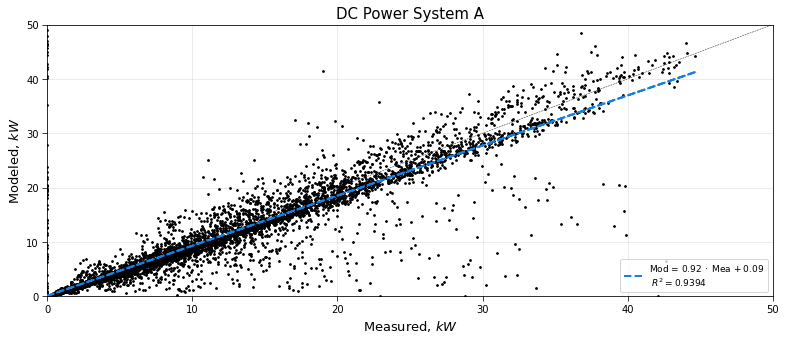}
   \label{fig:C2F10a} 
\end{subfigure}

\begin{subfigure}{1\textwidth}
   \centering
   \includegraphics[width=0.70\linewidth]{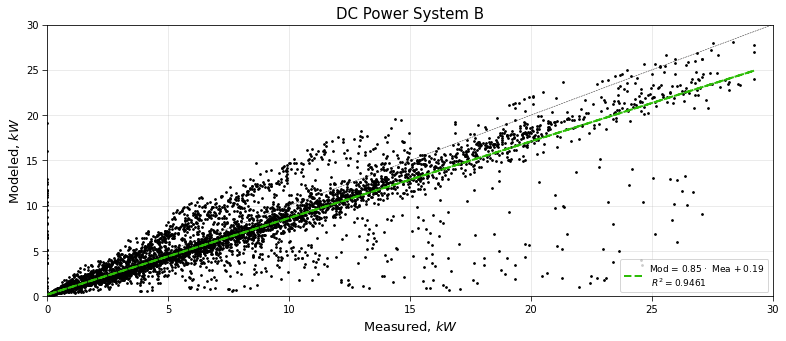}
   \label{fig:C2F10b} 
\end{subfigure}
\caption{Correlation between measured and modeled DC power for System A (top) and System B (bottom).}
\label{fig:C2F10}
\end{figure}

Figure \ref{fig:C2F10} helps understand the precision of the computational model. An exact simulation would have a slope and intercept equivalent to 1 and 0, respectively. The difference in slopes compared to ideal values is 6\% and 5.4\% for System A and System B, respectively.
AC power exhibits a behavior similar to that of DC power, as the statistical indicators are close to each other. However, the $R^2$ for System A is 2.3\% closer to the ideal value of $R^2$, and the MeAPE decreased. This improvement is attributed to the SNL model used to calculate AC power. Moreover, the SNL model considers inverter saturation limits, acting as a filter that restricts extreme values.

\begin{figure}[]
\centering
\begin{subfigure}{1\textwidth}
   \centering
   \includegraphics[width=0.70\linewidth]{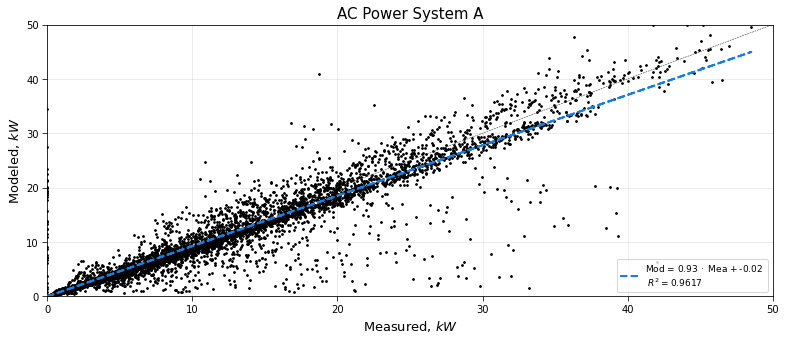}
   \label{fig:C2F11a} 
\end{subfigure}

\begin{subfigure}{1\textwidth}
   \centering
   \includegraphics[width=0.70\linewidth]{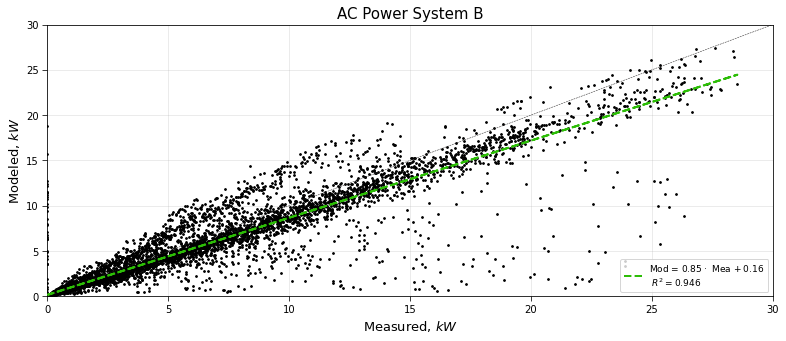}
   \label{fig:C2F11b} 
\end{subfigure}
\caption{Correlation between measured and modeled AC power for System A (top) and System B (bottom).}
\label{fig:C2F11}
\end{figure}

\subsubsection{Energy}
\label{subsec:C2S3S5}
\begin{figure}[]
\centering
\includegraphics[width=0.70\textwidth]{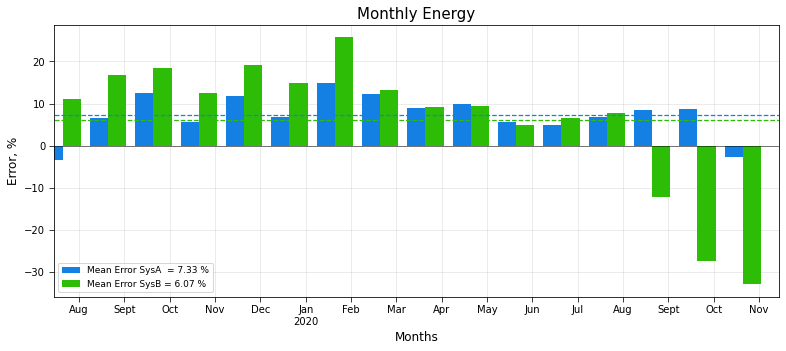}
\caption{Percentage error of the simulated energy with respect to the measurements. Blue bars correspond to System A and green bars to System B.}
\label{fig:C2F12}
\end{figure}

Figure \ref{fig:C2F12} illustrates the percentage error of simulated energy compared to measurement.

The variability in percentage errors is attributed to the dynamic meteorological conditions, which can be challenging to computationally model. However, the mean error amounts to approximately 7.3\% and 6.1\% for System A and System B, respectively. This mean error value is close to the ideal range of $\leq 5\%$. Additionally, the mean RMSE for both systems is 0.4 MWh. 

On the other hand, the $R^2$ indicates a good correlation between simulation and measurement for System A. The value for System B deviates by approximately 12 percentage points. This discrepancy is due to System B failures from Sept- to Nov-2020. The failure is associated with some strings of System B. The $R^2$ for System B is 0.95 upon recalculating  excluding these months.

Finally, the MeAPE of both systems are close to each other, indicating the absence of extreme data points in the dataset that could bias the analysis. Additionally, the MeAPE for System A (7.3\%) is near the ideal range, indicating the model's accuracy. For System B, the MeAPE is 12.3\%, and its deviation from the expected range correlates with the challenges faced by this system in the latter months of 2020. The minimum and maximum APE values for System A and System B are 14.1\% and 1.9\%, and 31.8\% and 5.0\%, respectively.

\begin{table}[]
\centering
\caption{Accuracy metrics between measurement and modeling of the PV system simulation.}
\label{tab:C2T6}
\resizebox{\textwidth}{!}{%
\begin{tabular}{ccccccccccc}
\hline
 & \multicolumn{2}{c}{DC Voltage} & \multicolumn{2}{c}{DC Current} & \multicolumn{2}{c}{DC Power} & \multicolumn{2}{c}{AC Power} & \multicolumn{2}{c}{Monthly Energy} \\ \cline{2-11} 
 & System A & System B & System A & System B & System A & System B & System A & System B & System A & System B \\ \hline
RMSE & 66.8 & 74.6 & 5.5 & 3.2 & 3.8 & 2.2 & 3.3 & 2.2 & 0.2 & 0.6 \\
R$^2$ & 0.95 & 0.95 & 0.92 & 0.90 & 0.93 & 0.94 & 0.96 & 0.94 & 0.97 & 0.85 \\
MeAPE & 1.9 & 2.1 & 12.1 & 22.4 & 13.2 & 21.1 & 13.8 & 19.6 & 7.3 & 12.3 \\ \hline
\end{tabular}%
}
\end{table}

\subsubsection{Comparison between PVlib, PVWatts, and PVsyst Models}
\label{subsec:C2S3S6}
To compare the three main computational models, the daily energy produced during the third week of June 2020 is evaluated using the MAPE statistical indicator. In addition to using the same operational input data, the algorithms include an equal percentage of losses to standardize the simulation. \cite{ref:C2R35} stipulates that under NOCT conditions, typically the percentage loss is 26.9\% of the installed capacity of the PV system.

All three models demonstrate acceptable precision in estimating daily energy, with MAPE falling within the ideal range ($\leq 5\%$). However, when comparing the mean MAPE between System A and System B, PVlib shows the lowest value (0.6\%), followed by PVWatts (1.8\%) and PVsyst (2.4\%).

This analysis highlights the competitiveness of PVlib against the most commonly used commercial models. Moreover, due to its flexibility and freedom in algorithm design and mathematical methods, PVlib allows for improvements in computational model accuracy.

\begin{figure}[]
\centering
\begin{subfigure}{0.475\textwidth}
   \centering
   \includegraphics[width=1\linewidth]{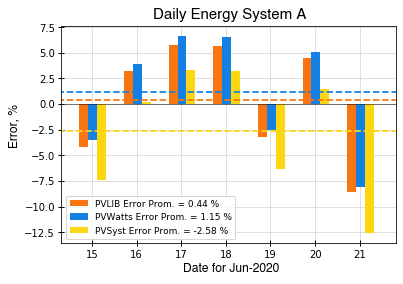}
   \label{fig:C2F13a} 
\end{subfigure}
\begin{subfigure}{0.465\textwidth}
   \centering
   \includegraphics[width=1\linewidth]{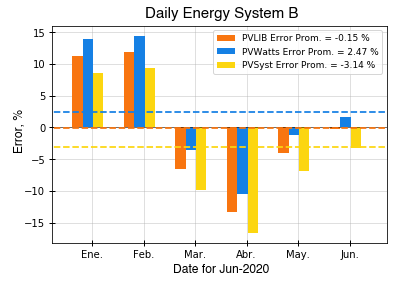}
   \label{fig:C2F13b} 
\end{subfigure}
\caption{Percentage error of the simulated energy with respect to the measurements. Orange represents PVlib, blue represents PVWatts, and yellow represents PVsyst.}
\label{fig:C2F13}
\end{figure}

\subsection{Synthetic Temperature}
\label{subsec:C3S3S3}

The synthetic datasets of $T_{amb}$ and $T_{cell}$ are compared with measurements reported by the monitoring system for the date Jan-2021. The results demonstrate the correlation between the generated data and the historical data for that date. It is observed that the synthetic behavior maintains physical consistency and follows the meteorological dynamics of the real data.

The synthetic $T_{cell}$ is also compared with the estimated by the NOCT model (Eq. \ref{eq:C3E1}). From Figure \ref{fig:C3F13}, it is observed that the NOCT model fails to capture the real taxonomy of the data as it only depends on $G_{POA}$ as a dynamic parameter, whereas the synthetic methodology successfully captures climatic variability.

\begin{equation}
\label{eq:C3E1}
T_{\text{cell}} = T_{\text{amb}} + \left( \frac{T_{\text{NOCT}} - 20}{800} \right) G_{\text{POA}} \;
\end{equation}

Comparing the parameters of a linear regression for synthetic $T_{\text{cell}}$ and the measurement with respect to $G_{POA}$, we find the same slope of magnitude 0.04. The intercept differs by 0.14$^\circ$C (12.3$^\circ$C and 12.4$^\circ$C, respectively), and the $R^2$ varies by approximately 0.3\% (0.942 and 0.945, respectively).

\begin{figure}[]
\centering
\begin{subfigure}{0.475\textwidth}
   \centering
   \includegraphics[width=1\linewidth]{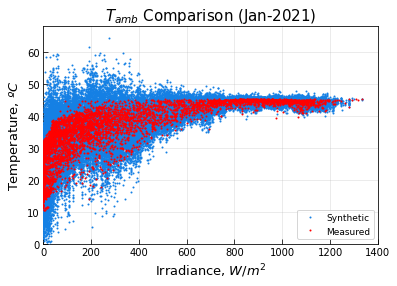}
   \label{fig:C3F13a} 
\end{subfigure}
\begin{subfigure}{0.465\textwidth}
   \centering
   \includegraphics[width=1\linewidth]{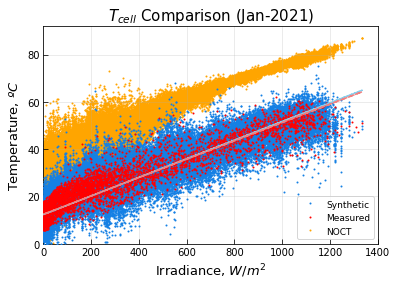}
   \label{fig:C3F13b} 
\end{subfigure}

\caption{Comparison between synthetic temperature datasets with measurements.}
\label{fig:C3F13}
\end{figure}

\subsection{Statistical Metrics of the ANN}
\label{subsec:C3S3S4}

\subsubsection{Photoluminescence Current $I_L$}
\label{subsec:C3S3S4S1}
From the training and validation curves (see Figure \ref{fig:C3F13}), it is observed that with the initial LR of 0.1, the training curve closely follows the validation curve shortly after training begins. By the third epoch, a plateau in the network's loss function indicates convergence to a local minimum. At this point, the LR is adjusted to attempt to find another local minimum, as evidenced by the step changes. For the fourth fold, adjusting the LR led to a local minimum with poorer performance compared to the other folds. The closest proximity between the training and validation curves occurs towards the end of training, specifically after 38 epochs, confirming the convergence and good performance of the ANN.

The $R^2$ (see Table \ref{tab:C3T4}) demonstrates a positive correlation between prediction and target value across all folds, indicating that the ANN's performance is representative and independent of the dataset. Furthermore, the minimum reported MeAPE for System A is 13.3\%.

The ANN architecture for System B's dataset is identical to that of System A. The metrics obtained are very similar to the results for System A (see Table \ref{tab:C3T4}), indicating the robustness of the prediction algorithm and its capability to perform across PV systems of different installed capacities; the mean difference between indicators for each system is approximately 2\%.

\begin{figure}[]
\centering
\begin{subfigure}{0.25\textwidth}
   \centering
   \includegraphics[width=1\linewidth]{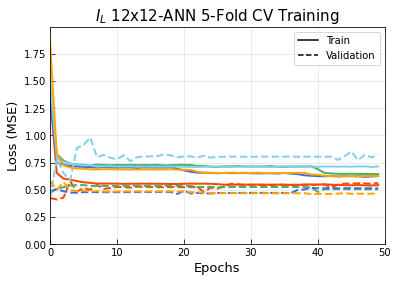}
   \label{fig:C3F13a} 
\end{subfigure}
\begin{subfigure}{0.7\textwidth}
   \centering
   \includegraphics[width=1\linewidth]{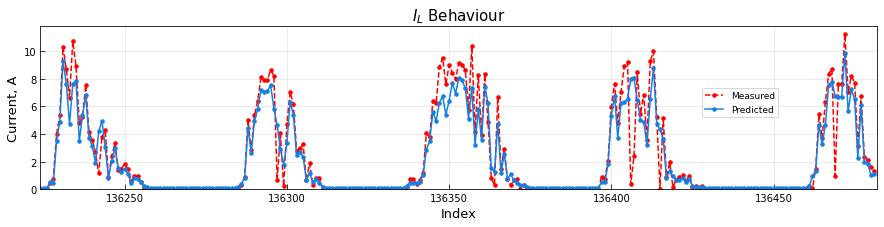}
   \label{fig:C3F13b} 
\end{subfigure}

\caption{Evolution of the loss function over epochs (left), and comparison of measured and modeled behavior of parameter $I_{L}$ (right). Results of System A.} 
\label{fig:C3F13}
\end{figure}

\subsubsection{Short-Circuit Current $I_{sc}$}
\label{subsec:C3S3S4S2}
\begin{figure}[]
\centering
\begin{subfigure}{0.25\textwidth}
   \centering
   \includegraphics[width=1\linewidth]{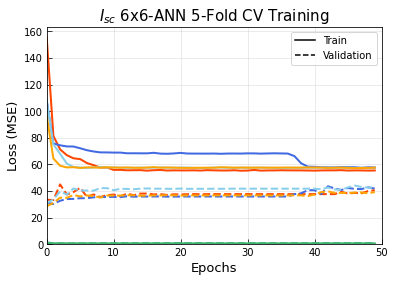}
   \label{fig:C3F14a} 
\end{subfigure}
\begin{subfigure}{0.7\textwidth}
   \centering
   \includegraphics[width=1\linewidth]{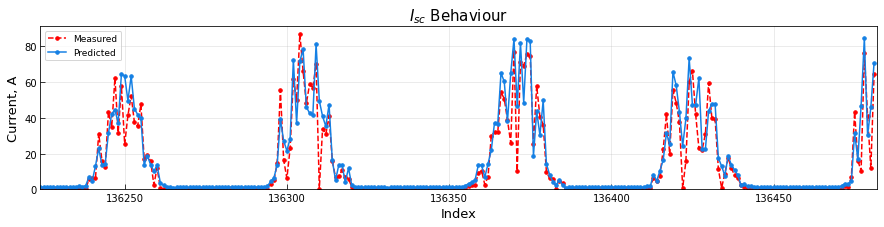}
   \label{fig:C3F14c} 
\end{subfigure}

\caption{Evolution of the loss function over epochs (left), and comparison of measured and modeled behavior of parameter $I_{sc}$ (right). Results of System A.} 
\label{fig:C3F14}
\end{figure}

The training curves (seee Figure \ref{fig:C3F14}) show a rapid convergence of the loss function from the beginning. The fact that all validation curves are lower than those of training is due to the implementation of dropout, which prevents overfitting when tuning the network's hyperparameters but is not active during validation.

Furthermore, similar to $I_{L}$, all folds exhibit a $R^2$ indicating a positive correlation close to the ideal. The RMSE indicates an mean prediction error of 4.9\%.

The fact that MeAPE is around 23\% suggests that there are instances where the ANN's prediction deviates from the target value. Examining the comparison curves, moments of sharp decline in the real value are evident which are not followed by the simulation; this indicates that the network fails to fully represent the stochastic nature of system operation failure occurrences. In the best scenario, the minimum MeAPE corresponds to 19.2\%.

The architecture of System B is identical to that of System A (see Table \ref{tab:C3T5}). However, the results show an 88.6\% improvement in the RMSE, indicating a lower mean error and thus a more accurate prediction. However, the MeAPE is higher by about 6\%, indicating that the error largely stems from the butterfly effect. The $R^2$ shows a 2\% improvement, indicating a better correlation fit for System B's dataset. These differences show that the proposed architecture manages to capture more precise patterns for System B's information. Jowever, the performance demonstrates that the ANN adjusts well to the technical and operational parameters of the PV sysetm.

\subsubsection{Open-Circuit Voltage $V_{oc}$}
\label{subsec:C3S3S4S3}
\begin{figure}[]
\centering
\begin{subfigure}{0.25\textwidth}
   \centering
   \includegraphics[width=1\linewidth]{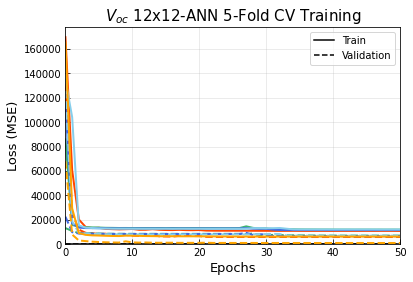}
   \label{fig:C3F15a} 
\end{subfigure}
\begin{subfigure}{0.7\textwidth}
   \centering
   \includegraphics[width=1\linewidth]{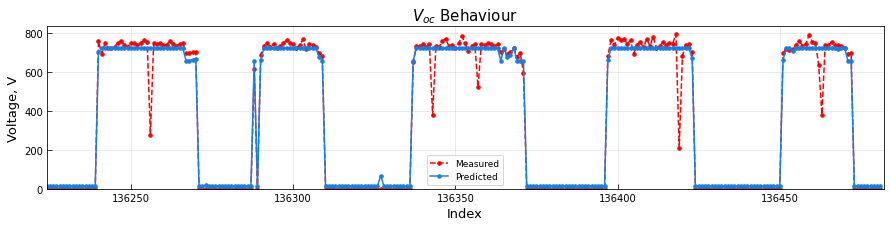}
   \label{fig:C3F15c} 
\end{subfigure}

\caption{Evolution of the loss function over epochs (left), and comparison of measured and modeled behavior of parameter $V_{oc}$ (right). Results of System A.} 
\label{fig:C3F15}
\end{figure}

The training and validation curves (see Figure \ref{fig:C3F15}) indicate rapid convergence of the loss function. Additionally, with the initial LR and its subsequent adjustment using the ReduceLROnPlateau scheduler, the two curves approach each other as epochs progress.

The $R^2$ value of 0.96 indicates a strong positive correlation between prediction and measurement. The fact that all folds are close to each other indicates a representative performance of the network independent of the dataset. 

Moreover, a stratified dispersion is observed in values around 500, 750, and 700. This is because the signal of $V_{oc}$ parameter behaves similarly to a square wave, and faults involve a reduction in its amplitude. Therefore, this range of values is more challenging for the ANN to predict accurately, as its patterns primarily recognize nominal voltage values. However, the ANN manages to provide a good fit for most of the fault instances. Overall, the ANN demonstrates acceptable accuracy, with an RMSE equivalent to approximately 10\% of the nominal voltage.

The ANN achieves an adjustment with accuracy within the ideal range according to the MeAPE value (3.9\%). This precision is observed in the comparison of the $V_{oc}$ signal generated by the ANN and the corresponding expected value.

System B, whose ANN architecture is equivalent to that of System A (see Table \ref{tab:C3T5}), shows greater accuracy in data fitting, as MeAPE is 27.9\% lower than System A.

\subsubsection{PV Panel Efficiency $\eta_{cell}$}
\label{subsec:C3S3S4S4}
\begin{figure}[]
\centering
\begin{subfigure}{0.25\textwidth}
   \centering
   \includegraphics[width=1\linewidth]{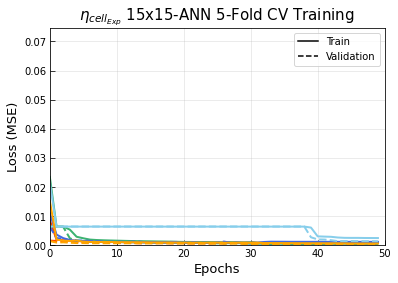}
   \label{fig:C3F16a} 
\end{subfigure}
\begin{subfigure}{0.7\textwidth}
   \centering
   \includegraphics[width=1\linewidth]{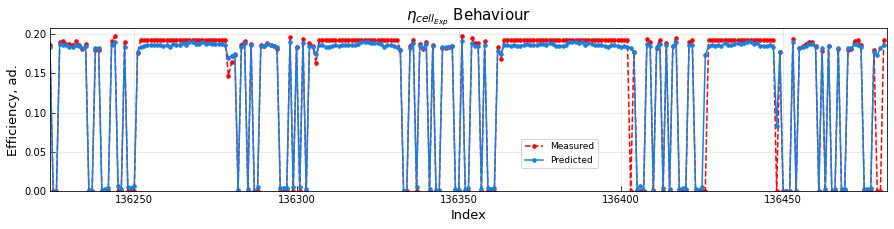}
   \label{fig:C3F16c} 
\end{subfigure}

\caption{Evolution of the loss function over epochs (left), and comparison of measured and modeled behavior of parameter $\eta_{cell}$ (right). Results of System A.} 
\label{fig:C3F16}
\end{figure}

The training and validation curves (see Figure \ref{fig:C3F16}) show rapid convergence except for the fourth fold. The weight initialization for this fold localized the learning process in a local minimum of different magnitude compared to the others. However, with the adaptive LR strategy using the ReduceLROnPlateu scheduler, around epoch 40, an LR adjustment is made, as visualized by the step in the light blue curve, resulting in improved model performance from this point onward.

The correlation between prediction and target value shows that the ANN manages to capture the behavior of high efficiencies of the PV panel and their variability due to fluctuations in production or included faults. At low efficiency magnitudes the ANN fails to identify sufficient patterns to generate high-precision predictions. This is due to the erratic behavior of the PV panel efficiency once losses are included, as eight out of eleven faults considered directly affect the power produced by the PV panel. Nevertheless, the $R^2$ equivalent to 0.89 indicates that the ANN identifies the trend in the $\eta_{cell}$ taxonomy.

The RMSE is 15\% of the maximum measured value. The dispersion of the prediction at low magnitudes leads to this percentage, as it is observed that the values of $\eta_{cell}$ are mainly concentrated at the extremes and that the ANN does capture the mean behavior at high magnitudes.

The mean MeAPE for the dataset (4.9\%) is within the ideal range (i.e., $\leq$ 5\%), which validates the overall accuracy of the ANN. The performance of the ANN is observed by detailing the comparison of the signal generated by the ANN and the target value, where efficiency fluctuations due to the inclusion of faults in $P_{DC}$ are identified and correctly predicted.

The ANN architecture designed for the System B dataset is identical to System A (see Table \ref{tab:C3T5}). The statistical metrics show an improvement compared to the performance of System A, despite having an increase in RMSE. The 6.7\% increase in the $R^2$ indicates a stronger correlation between the prediction and the measured value of the dataset. The MeAPE validates the accuracy of the model by not taking into account discrepancies caused by outliers.

\subsubsection{Inverter Efficiency $\eta_{inv}$}
\label{subsec:C3S3S4S5}
\begin{figure}[]
\centering
\begin{subfigure}{0.25\textwidth}
   \centering
   \includegraphics[width=1\linewidth]{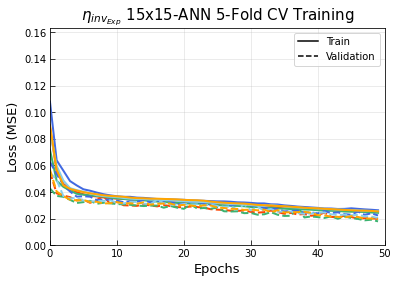}
   \label{fig:C3F17a} 
\end{subfigure}
\begin{subfigure}{0.7\textwidth}
   \centering
   \includegraphics[width=1\linewidth]{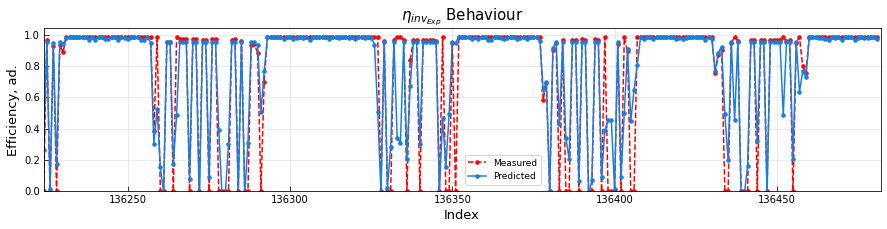}
   \label{fig:C3F17c} 
\end{subfigure}

\caption{Evolution of the loss function over epochs (left), and comparison of measured and modeled behavior of parameter $\eta_{inv}$ (right). Results of System A.} 
\label{fig:C3F17}
\end{figure}

The training and validation curves (see Figure \ref{fig:C3F17}) indicate accuracy in predictions due to the proximity of the curves. However, no plateau is visualized, therefore more pochs are required for total convergence in training. A slightly stepped behavior is observed as the loss function reduces; this corresponds to the adaptive LR strategy.

The correlation shows that in all folds, the ANN makes the best predictions at high efficiency magnitudes. This demonstrates that the ANN captures the patterns describing the variability in this data range, which appear closer to each other. On the other hand, for magnitudes below this value, the prediction dispersion is greater. This is due to the erratic behavior induced by the faults at these moments. However, the RMSE indicates that the mean prediction error for System A and System B is 0.14 (see Table \ref{tab:C3T4}). The $R^2$ shows a positive correlation between the prediction and the target value, although the magnitudes obtained for both System A and System B are the lowest compared to the other study parameters (0.8 and 0.7, respectively).

The MeAPE is 77.8\% lower for both System A and System B, indicating a more accurate prediction of the ANN.

\begin{table}[]
\centering
\caption{Five-fold cross-validation accuracy metrics for ANN output parameter prediction.}
\label{tab:C3T4}
\resizebox{\textwidth}{!}{%
\begin{tabular}{ccccccccccc}
\hline
 & \multicolumn{2}{c}{$I_L$} & \multicolumn{2}{c}{$I_{sc}$} & \multicolumn{2}{c}{$V_{oc}$} & \multicolumn{2}{c}{$\eta_{cell}$} & \multicolumn{2}{c}{$\eta_{inv}$} \\ \cline{2-11} 
 & System A & System B & System A & System B & System A & System B & System A & System B & System A & System B \\ \hline
RMSE & 0.6 & 0.5 & 4.9 & 0.5 & 70.7 & 88.5 & 0.03 & 0.5 & 0.1 & 0.1 \\
R$^2$ & 0.94 & 0.95 & 0.93 & 0.95 & 0.96 & 0.95 & 0.89 & 0.95 & 0.80 & 0.77 \\
MeAPE & 17.4 & 18.6 & 23.6 & 29.3 & 3.9 & 2.8 & 4.98 & 3.6 & 1.17 & 0.58 \\ \hline
\end{tabular}%
}
\end{table}

\begin{table}[]
\centering
\caption{ANN architecture for each output parameter.}
\label{tab:C3T5}
\begin{tabular}{cccccc}
\hline
ANN Architecture & $I_L$ & $I_{sc}$ & $V_{oc}$ & $\eta_{cell}$ & $\eta_{inv}$ \\ \hline
Hidden Layers & 2 & 2 & 2 & 2 & 2 \\
Neurons per Layer & 12 & 6 & 12 & 15 & 15 \\ \hline
\end{tabular}
\end{table}

\subsection{Computational Simulation Validation}
\label{sec:C4S3}
\begin{figure}[]
\centering
\begin{subfigure}{0.475\textwidth}
   \centering
   \includegraphics[width=0.8\linewidth]{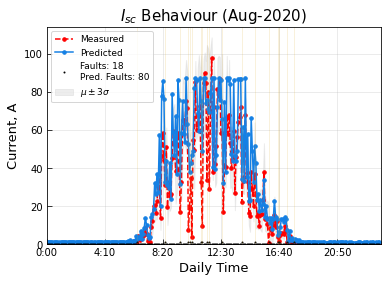}
   \label{fig:C4F2a} 
\end{subfigure}
\begin{subfigure}{0.465\textwidth}
   \centering
   \includegraphics[width=0.8\linewidth]{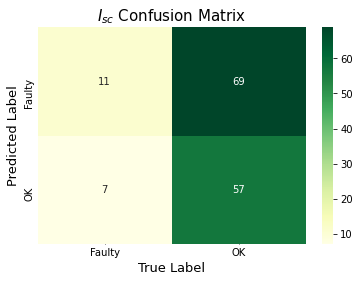}
   \label{fig:C4F2b} 
\end{subfigure}

\begin{subfigure}{0.475\textwidth}
   \centering
   \includegraphics[width=0.8\linewidth]{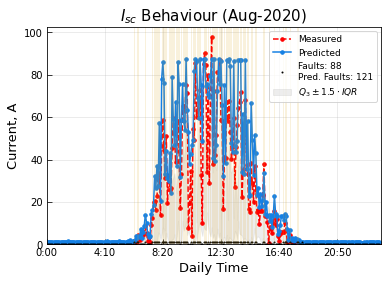}
   \label{fig:C4F2c} 
\end{subfigure}
\begin{subfigure}{0.465\textwidth}
   \centering
   \includegraphics[width=0.8\linewidth]{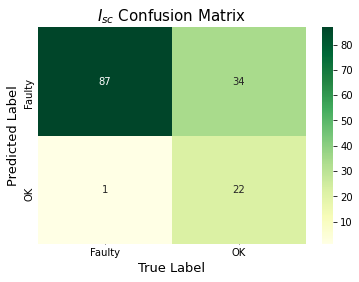}
   \label{fig:C4F2d} 
\end{subfigure}

\begin{subfigure}{0.475\textwidth}
   \centering
   \includegraphics[width=0.8\linewidth]{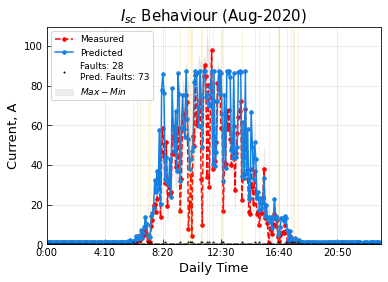}
   \label{fig:C4F1e} 
\end{subfigure}
\begin{subfigure}{0.465\textwidth}
   \centering
   \includegraphics[width=0.8\linewidth]{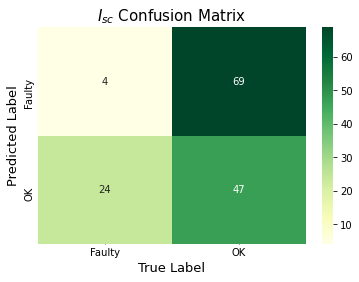}
   \label{fig:C4F2f} 
\end{subfigure}
\caption{Prediction and measurement of $I_{sc}$ on the left, and the corresponding confusion matrix according to the threshold. The limits are defined by $\mu \pm 3\sigma$ (top), $Q_3 \pm 1.5 IQR$ (middle) and Maximum and Minimum (bottom).}
\label{fig:C4F2}
\end{figure}

\begin{figure}[]
\centering
\begin{subfigure}{0.475\textwidth}
   \centering
   \includegraphics[width=0.8\linewidth]{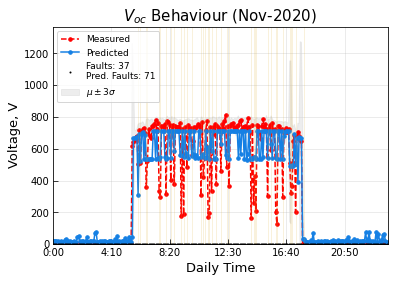}
   \label{fig:C4F3a} 
\end{subfigure}
\begin{subfigure}{0.465\textwidth}
   \centering
   \includegraphics[width=0.8\linewidth]{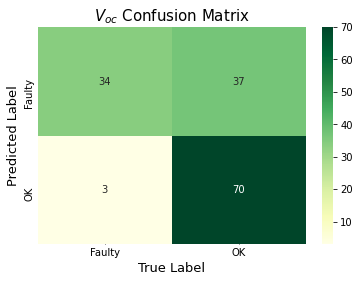}
   \label{fig:C4F3b} 
\end{subfigure}

\begin{subfigure}{0.475\textwidth}
   \centering
   \includegraphics[width=0.8\linewidth]{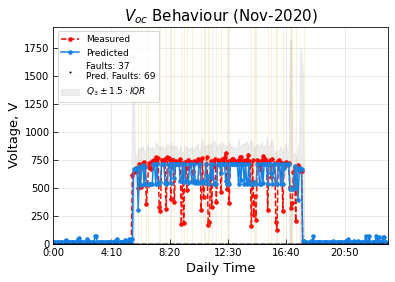}
   \label{fig:C4F3c} 
\end{subfigure}
\begin{subfigure}{0.465\textwidth}
   \centering
   \includegraphics[width=0.8\linewidth]{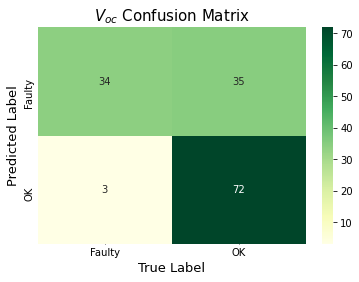}
   \label{fig:C4F3d} 
\end{subfigure}

\begin{subfigure}{0.475\textwidth}
   \centering
   \includegraphics[width=0.8\linewidth]{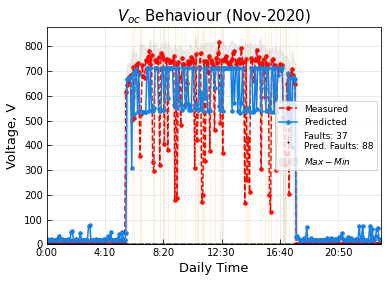}
   \label{fig:C4F3e} 
\end{subfigure}
\begin{subfigure}{0.465\textwidth}
   \centering
   \includegraphics[width=0.8\linewidth]{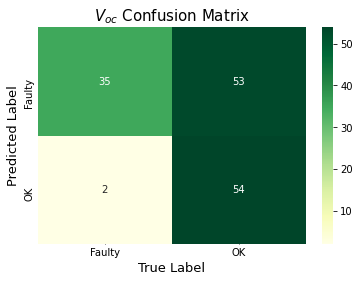}
   \label{fig:C4F3f} 
\end{subfigure}
\caption{Prediction and measurement of $V_{oc}$ on the left, and corresponding confusion matrix according to the threshold. The thresholds are defined by $\mu \pm 3\sigma$ (top), $Q_3 \pm 1.5 IQR$ (middle), and Maximum and Minimum (bottom).}
\label{fig:C4F3}
\end{figure}

From Figure \ref{fig:C4F1} to \ref{fig:C4F5}, the signals predicted by the ANN and the corresponding target value are shown. The gray background represents the normal operating range defined by the thresholds obtained for each studied alternative (recall Section \ref{subsec:C4S2S1}). When the predicted signal violates the limits, a fault is detected at that moment; otherwise, the operation is within the normal range. The detected faults are visually supported by the yellow vertical lines. The respective confusion matrix are also presented to quantify the accuracy in fault detection.

\subsubsection{Photoluminescence Current $I_L$}
\label{subsec:C3S3S1}
\begin{figure}[]
\centering
\begin{subfigure}{0.475\textwidth}
   \centering
   \includegraphics[width=0.8\linewidth]{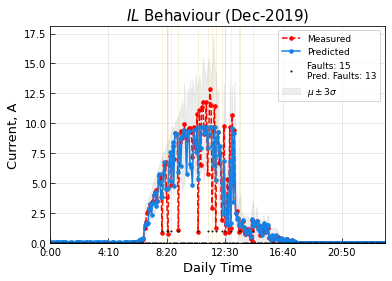}
   \label{fig:C4F1a} 
\end{subfigure}
\begin{subfigure}{0.465\textwidth}
   \centering
   \includegraphics[width=0.8\linewidth]{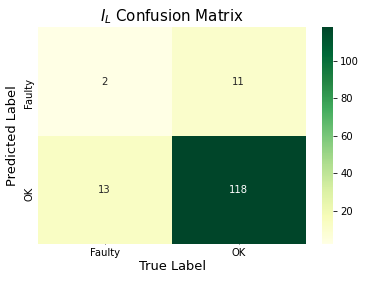}
   \label{fig:C4F1b} 
\end{subfigure}

\begin{subfigure}{0.475\textwidth}
   \centering
   \includegraphics[width=0.8\linewidth]{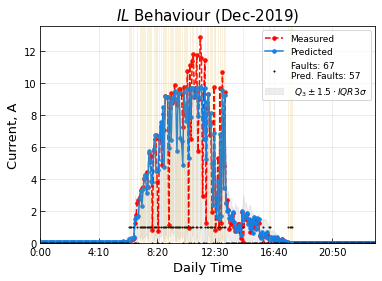}
   \label{fig:C4F1c} 
\end{subfigure}
\begin{subfigure}{0.465\textwidth}
   \centering
   \includegraphics[width=0.8\linewidth]{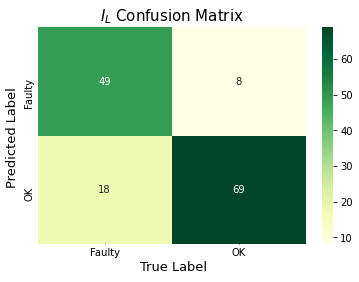}
   \label{fig:C4F1d} 
\end{subfigure}

\begin{subfigure}{0.475\textwidth}
   \centering
   \includegraphics[width=0.8\linewidth]{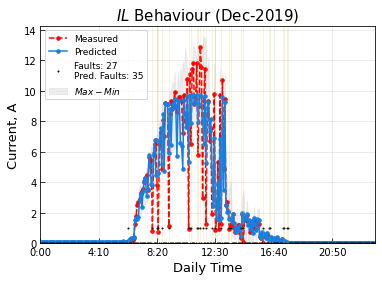}
   \label{fig:C4F1e} 
\end{subfigure}
\begin{subfigure}{0.465\textwidth}
   \centering
   \includegraphics[width=0.8\linewidth]{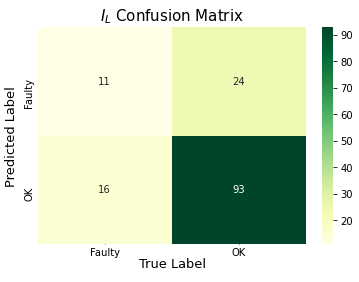}
   \label{fig:C4F1f} 
\end{subfigure}
\caption{Prediction and measurement of $I_{L}$ on the left, and corresponding confusion matrix according to the threshold. The limits are defined by $\mu \pm 3\sigma$ (top), $Q_3 \pm 1.5 IQR$ (middle), and Maximum and Minimum (bottom).}
\label{fig:C4F1}
\end{figure}

The ANN fails to capture the incorporated low-magnitude losses. Therefore, the definition of thresholds is applied to the target values in order to extract the high-magnitude losses added to the dataset and evaluate the accuracy in their detection (see Figure \ref{fig:C4F1}).

The results of the confusion matrix show similar accuracy for the $\mu \pm 3\sigma$ and $Q_3 \pm 1.5 IQR$ alternatives, with a mean accuracy of 80.8\% and 78.7\%, respectively. On the other hand, the maximum and minimum alternative achieved a mean accuracy of 70.6\%.

As it is observed, the prominent fault peaks in the target value are not predicted with complete accuracy by the ANN; these values, which are truly faulty (i.e., 1), can be incorrectly classified as not faulty (i.e., 0).

\subsubsection{Short-Circuit Current $I_{sc}$}
\label{subsec:C3S3S2}
Similar to the parameter $I_L$, the ANN fails to detect low magnitude faults of the nominal maximum fault. Therefore, the thresholds of each alternative are applied to the target values and those detected as faults are compared with the corresponding prediction; this means that the ANN manages to detect high magnitude faults that significantly impact the operating signal. On the other hand, this behavior is consistent with respect to $I_L$ due to its mathematical dependence. This correlation is validated by detailing that the accuracy differs by 6.4\% with respect to the variable $I_L$.

On the other hand, the alternative $Q_3 \pm 1.5 IQR$ obtains the highest accuracy (75.6\%), while the others fail to capture the data patterns as they present an accuracy of 52.4 and 58.2\% for the limits given by $\mu \pm 3\sigma$ and maximum and minimum, respectively.

Both the accuracy of $I_L$ and $I_{sc}$ is affected by false alarms, that is, faulty detections but whose target value is in an non-faulty state. This indicates that the ANN manages to replicate the variability of the signal but in terms of general behavior.

\subsubsection{Open-Circuit Voltage $V_{oc}$}
\label{subsec:C4S3S3}
The results of the confusion matrix show similar accuracy for the $\mu \pm 3\sigma$ and maximum and minimum alternatives, with an accuracy of 77.5 and 72.0\%, respectively. The threshold definition that achieves the highest accuracy is from the $Q_3 \pm 1.5 IQR$ alternative, and a maximum accuracy of 98.7\% achieved.

The confusion matrices also show that the faulty data is correctly detected, with a maximum of three incorrect classifications for $Q_3 \pm 1.5 IQR$. However, many FP are evident which are interpreted as false alarms; this is caused because the ANN fails to detail all fluctuations in the target value.

\subsubsection{PV Panel Efficiency $\eta_{cell}$}
\label{subsec:C4S3S4}
The ANN successfully reproduces the behavior of PV panel efficiency (i.e., nominal value of 19.3\%, capturing fluctuations due to the inclusion of losses). This is demonstrated by the confusion matrix, where for all three alternatives defining normal operation thresholds, the mean accuracy is 91.6\% with a deviation of 1.4\%, specifically achieving accuracies of 91.6, 92.6, and 90.5 for the $\mu \pm 3\sigma$, $Q_3 \pm 1.5 IQR$, and maximum and minimum alternatives, respectively. Similar to the previous parameters, the alternative resulting in the highest accuracy is $Q_3 \pm 1.5 IQR$.

\begin{figure}[]
\centering
\begin{subfigure}{0.475\textwidth}
   \centering
   \includegraphics[width=0.8\linewidth]{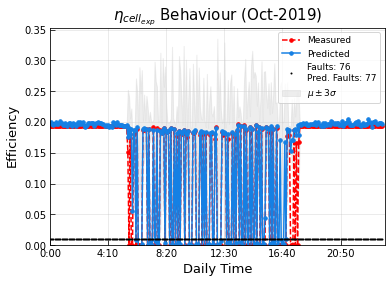}
   \label{fig:C4F4a} 
\end{subfigure}
\begin{subfigure}{0.465\textwidth}
   \centering
   \includegraphics[width=0.8\linewidth]{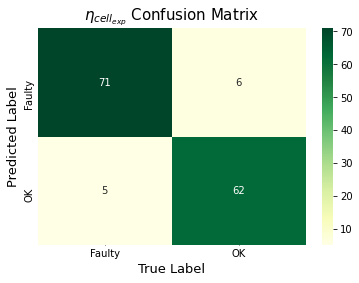}
   \label{fig:C4F4b} 
\end{subfigure}

\begin{subfigure}{0.475\textwidth}
   \centering
   \includegraphics[width=0.8\linewidth]{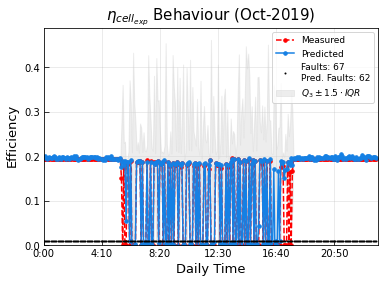}
   \label{fig:C4F4c} 
\end{subfigure}
\begin{subfigure}{0.465\textwidth}
   \centering
   \includegraphics[width=0.8\linewidth]{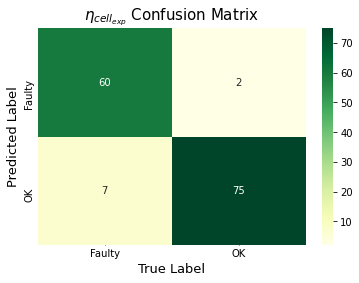}
   \label{fig:C4F4d} 
\end{subfigure}

\begin{subfigure}{0.475\textwidth}
   \centering
   \includegraphics[width=0.8\linewidth]{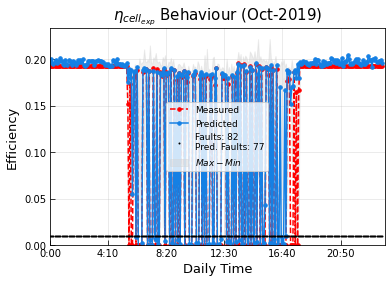}
   \label{fig:C4F4e} 
\end{subfigure}
\begin{subfigure}{0.465\textwidth}
   \centering
   \includegraphics[width=0.8\linewidth]{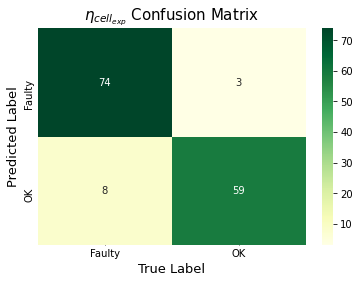}
   \label{fig:C4F4f} 
\end{subfigure}
\caption{Prediction and measurement of $\eta_{cell}$ on the left, and corresponding confusion matrix according to the threshold. The limits are defined by $\mu \pm 3\sigma$ (top), $Q_3 \pm 1.5 IQR$ (middle), and Maximum and Minimum (bottom).}
\label{fig:C4F4}
\end{figure}

Finally, for $\eta_{cell}$, few FP are observed, a result of pattern recognition by the ANN. Unlike previous parameters where errors are driven by false alarms, in this case, the ANN effectively identifies defective data as not faulty (i.e., 1). From Figure \ref{fig:C4F4}, it is observed that this occurs during sunrise and sunset times, where the ANN shows greater deviation from the target value.

\subsubsection{Inverter Efficiency $\eta_{inv}$}
\label{subsec:C4S3S5}
The definition of normal operation thresholds using $\mu \pm 3\sigma$ achieves the highest accuracy in fault detection (76.5\%), followed by the $Q_3 \pm 1.5 IQR$ alternative (75.6\%), and finally with maximum and minimum (72.1\%). The highest precision achieved is 92.3\% with the first alternative.

\begin{figure}[]
\centering
\begin{subfigure}{0.475\textwidth}
   \centering
   \includegraphics[width=0.8\linewidth]{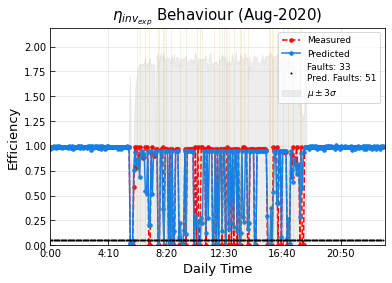}
   \label{fig:C4F5a} 
\end{subfigure}
\begin{subfigure}{0.465\textwidth}
   \centering
   \includegraphics[width=0.8\linewidth]{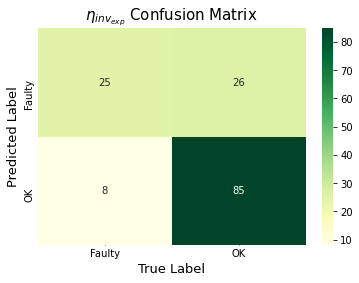}
   \label{fig:C4F5b} 
\end{subfigure}

\begin{subfigure}{0.475\textwidth}
   \centering
   \includegraphics[width=0.8\linewidth]{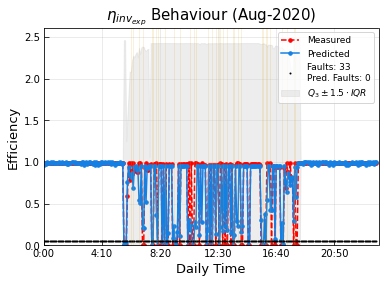}
   \label{fig:C4F5c} 
\end{subfigure}
\begin{subfigure}{0.465\textwidth}
   \centering
   \includegraphics[width=0.8\linewidth]{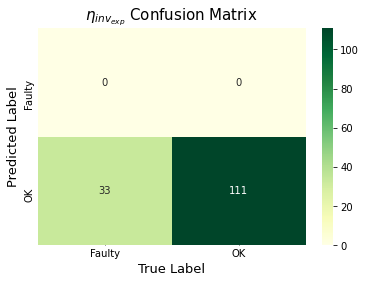}
   \label{fig:C4F5d} 
\end{subfigure}

\begin{subfigure}{0.475\textwidth}
   \centering
   \includegraphics[width=0.8\linewidth]{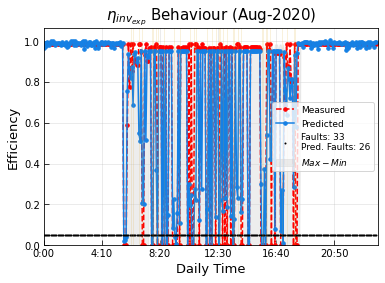}
   \label{fig:C4F5e} 
\end{subfigure}
\begin{subfigure}{0.465\textwidth}
   \centering
   \includegraphics[width=0.8\linewidth]{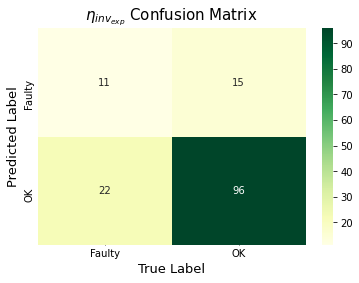}
   \label{fig:C4F5f} 
\end{subfigure}
\caption{Prediction and measurement of $\eta_{inv}$ on the left, and corresponding confusion matrix based on the threshold. The thresholds are defined by $\mu \pm 3\sigma$ (top), $Q_3 \pm 1.5 IQR$ (middle), and Maximum and Minimum (bottom).}
\label{fig:C4F5}
\end{figure}

From the confusion matrices (see Figure \ref{fig:C4F5}), it is observed that non-faulty states are mostly correctly classified. Additionally, in some scenarios, the ANN fails to capture efficiency fluctuations resulting in misclassification.

\subsubsection{Comparison with Literature}
\label{subsec:C4S3S6}
Five signals are studied: $I_{L}$, $I_{sc}$, $V_{oc}$, $\eta_{cell}$ and $\eta_{inv}$. The mean accuracy obtained in fault detection is 82.2\%.

\cite{ref:C4R7} estimate operational parameters of current, voltage, and number of peaks in the I-V curve using an ANN to identify faults such as short circuit, open circuit, bypass diode deviation, reversed bypass diode, connection fault, partial shading, shading effect with faulty bypass diode, and shading effect with connection fault in PV systems. The ANN results are compared with corresponding measurements to perform faulty operation detection. Their results achieve an accuracy of 68.4\% using a RBFNN and 90.3\% using an ANN.

Similarly, \cite{ref:C4R8} integrates the ANN with RBFNN to detect faults based on voltage and power variations, achieving an accuracy of 92.1\% in the best scenario with an ANN architecture consisting of two inputs, two hidden layers with seven neurons each, and nine outputs.

The model proposed by \cite{ref:C4R6} correlates irradiance with normal operating limits, defined by $\mu \pm 3\sigma$. Their proposal achieves an accuracy of approximately 81.1\% when evaluating thresholds over the entire irradiance dataset, whereas setting limits in intervals of 200 W/m$^2$ results in an accuracy of approximately 96.2\%.

The novelty of the proposed approach in this research compared to those mentioned lies in generating technical signals from equipment whose monitoring is non-trivial and performing fault detection based on the historical behavior of the parameter under study. This taxonomy is grounded in the correlation of weather conditions and their influence on the dynamics of energy production in PV systems.

\section{Conclusion}
\label{sec-7}

The computational model developed in PVlib based on the equivalent circuit proposed by \cite{ref:C2R4} and modified by the California Energy Commission, and the mathematical modeling developed by Sandia National Laboratories in \cite{ref:C2R17}, proves to be a reliable representation of the photovoltaic (PV) system under study. The main sources of error include external factors affecting production that the computational model fails to capture, such as shading of the irradiance sensor (i.e., reference cell). 

The dynamic quantification of losses based on meteorological and operational system information plays a significant role in the computational accuracy achieved. The magnitude of losses aligns with literature values and exhibits behavior consistent with periodic preventive maintenance conducted at the plant.

The computational model in PVlib demonstrates higher accuracy in system performance estimates compared to two major commercial tools, PVWatts and PVsyst. Additionally, it overcomes the primary limitations of these computational models: (i) allows for comparing performance results using different models and assumptions; (ii) enables modification, customization, and updating of modeling algorithms; (iii) facilitates seamless integration into external workflow pipelines; and (iv) provides access to intermediate modeling results and enables on-the-fly statistical analysis.

Some limitations in the development of the computational model arise from certain unmonitored meteorological parameters that could contribute to higher accuracy in performance estimation (e.g., relative humidity, wind speed, and precipitation). Additionally, sophisticated techniques for estimating degradation losses, such as the year-on-year method proposed in \cite{ref:C2R31, ref:C2R32, ref:C2R33}, were not feasible to employ due to the PV system having just over a year and a half of operational data.

Synthetic irradiance dataset successfully capture the dynamic nature of climate and its historical in situ behavior. Daily characteristics (e.g., peak maximum, dispersion, taxonomy) remain within the expected range according to the meteorological condition on a specific given date. 

Synthetic ambient temperature mirrors the real values' behavior according to its dependency on irradiance. This is also evident for synthetic cell temperature, with the caveat that physical guidelines are better represented compared to a known equation such as NOCT model, thus offering greater precision.

Based on the developed computational model and the Gaussian or lognormal distribution of meteorological information \cite{SalazarPena2023}, the synthetic datasets overcomes the following limitations reported in the literature concerning artificial intelligence (AI) algorithms: (i) incorporates dynamic behavior with physical meaning, allowing exploration of the impact on performance from transient cloud effects; (ii) five-minute resolution enhances detailed simulations by capturing the intermittent nature of solar irradiance; and (iii) the occurrence of faults and accuracy in data quality knowledge ensures the reliability of the training set and the AI algorithm.

The design and training of the AI agent, i.e., artificial neural network (ANN), demonstrate learning in the relationship between input and output parameters of the PV system, such as meteorological and electrical parameters for production, respectively.

The response signal provided by the ANN captures the normal or faulty state of the target value; fluctuations due to weather dynamics and operational conditions are evident and consistent with the expected signal.

Statistical metrics indicate the presence of outliers in the signal generated by the ANN. These data occur at the sunrise and sunset (i.e., 6:00 and 18:00, respectively), so they do not represent significant deviations in simulated production. However, they noticeably affect the indicators estimating the model's accuracy; using the 75th percentile or robust metrics for extreme values corrects the deviation.

Three methods for defining normal operation thresholds are studied, taking into account the dynamic meteorological nature and its influence on production, which can cause unexpected fluctuations. Additionally, the defined thresholds are reliable and representative concerning historical data by considering their statistical distribution, being insensitive to outliers, and taking into account the usual maximum and minimum range for a specific date.

The normal operation threshold defined as $Q_3 \pm 1.5 IQR$, where $Q_3$ is the 75th percentile and $IQR$ is the interquartile range, exhibits the highest accuracy in fault detection due to its robustness against extreme values, thereby generating the least number of false alarms.

The mean accuracy achieved in fault detection is competitive with those reported in the literature. However, the novelty of this project lies in generating technical signals from equipment whose monitoring is not trivial and performing fault detection based on their historical behavior, considering the correlation with weather conditions and their influence on the dynamics of energy production in the PV system.

This research aims to contribute to the development of a PV solar energy framework both for academia and industry. The simulations validate the accurate computational modeling of the PV system located at Universidad de los Andes. The loss quantification methodologies assist in optimizing operational estimates of the PV system and seek to improve understanding of its operation and fluctuations due to weather conditions.

Furthermore, the computational model ensures reliable performance assessment of PV systems, enabling informed decision-making aimed at achieving expected efficiency. Loss estimation provides insight into the PV system's condition, facilitating project viability assessment by estimating revenue through energy yield estimation. Additionally, the development of a fault detection algorithm based on the PV system production data benefits the industry, as energy generation facilities typically have monitoring systems. This research aims to optimize PV system production by effectively detecting faults in monitored signals.
    
Finally, the methodology used to construct synthetic databases is valuable for various analyses in both academia and industry, as gathering training data through monitoring, especially fault occurrences, is not trivial nor straightforward. The proposed methodology considers meteorological dynamics and their impact on PV system performance, as well as the historical behavior.

\section{Future Work}
\label{sec-8}

\begin{enumerate}
    \item Implement the YOY algorithm to estimate degradation.
    
    \item Integrate the computational model online with the Meteocontrol monitoring system to enable real-time data acquisition and automated simulation.

    \item Develop a fault detection agent using a technique other than ANNs. Other techniques may have competitive accuracy compared to ANN, with the advantage of being more efficient and requiring lower computational cost.
    
    \item Incorporate faults into the database as a time series of the fault instead of producing data for instantaneous points.
    
    \item Evaluate the online implementation of the computational model and the fault detection agent for the evaluation of production information, since fault detection is based on synthetic data sets since their fault profile characteristics are known.
\end{enumerate}

\bibliographystyle{elsarticle-num}
\bibliography{references}

\end{document}